\documentclass[twocolumn,twocolappendix]{aastex631}
\usepackage{amsmath,mathtools,mathrsfs}
\usepackage{graphicx}
\usepackage{dcolumn}
\usepackage{bm}
\usepackage{accents}

\usepackage{ulem}
\usepackage{xcolor}

\DeclareFontFamily{OT1}{pzc}{}
\DeclareFontShape{OT1}{pzc}{m}{it}{<-> s * [1.10] pzcmi7t}{}
\DeclareMathAlphabet{\mathpzc}{OT1}{pzc}{m}{it}

\begin{document}

\title{Black Hole Images as Tests of General Relativity: Effects of Plasma Physics}

\author[0000-0003-4413-1523]{Feryal \"Ozel}
\affiliation{Department of Astronomy and Steward Observatory, University of Arizona, 933 N. Cherry Ave., Tucson, AZ 85721, USA}

\author[0000-0003-1035-3240]{Dimitrios Psaltis}
\affiliation{Department of Astronomy and Steward Observatory, University of Arizona, 933 N. Cherry Ave., Tucson, AZ 85721, USA}

\author[0000-0001-9283-1191]{Ziri Younsi}
\affiliation{Mullard Space Science Laboratory, University College London, Holmbury St.~Mary, Dorking, Surrey, RH5 6NT, UK}

\begin{abstract}
The horizon-scale images of black holes obtained with the Event Horizon Telescope have provided new probes of their metrics and tests of General Relativity. The images are characterized by a bright, near circular ring from the gravitationally lensed emission from the hot plasma and a deep central depression cast by the black hole. The metric tests rely on fact that the bright ring closely traces the boundary of the black hole shadow with a small displacement that has been quantified using simulations. In this paper we develop a self-consistent covariant analytic model of the accretion flow that spans a broad range of plasma conditions and black-hole properties to explore the general validity of this result. We show that, for any physical model of the accretion flow, the ring always encompasses the outline of the shadow and is not displaced by it by more than half the ring width. This result is a consequence of conservation laws and basic thermodynamic considerations and does not depend on the microphysics of the plasma or the details of the numerical simulations. We also present a quantitative measurement of the bias between the bright ring and the shadow radius based on the analytical models. 

\end{abstract}


\section{Introduction}

Horizon scale images of accreting black holes are generated when the photons emitted by the surrounding plasma propagate from the deep gravitational fields of the black holes to observers at infinity. Because of this, the strong fields of black holes are imprinted on the resulting images, which can be used to probe the spacetime properties.  

When observed at millimeter wavelengths, the radiatively inefficient accretion flows that surround nearby supermassive black holes are transparent down to the event horizon~\citep{Ozel2000}, allowing us to observe directly the photons that originate at horizon scales. These flows give rise to images in which the black holes cast a deep shadow on the plasma emission~\citep{Jaroszynski1997,Falcke2000}. 

The boundary of the shadow is determined entirely by the black-hole metric and, in particular, by the location of spherical photon orbits outside its horizon~\citep{Bardeen1973,Takahashi2004}. For a Kerr black hole, the shape of the shadow remains nearly circular and its size nearly constant for all black-hole spins and observer inclinations, as a result of a near cancellation of frame-dragging and spacetime-quadrupole effects~\citep{Johannsen2010b}. This property allows using black-hole shadows to perform a direct test of gravity at horizon scales that depends only on the prior knowledge of the black mass (see~\citealt{Psaltis2019} for a review).   

The Event Horizon Telescope (EHT) has recently obtained images of the black hole in the center of the M87 galaxy. These images are characterized by a narrow ring of emission that surrounds a deep brightness depression~\citep{PaperI,PaperIV}. They have been used to infer the size of the black-hole shadow~\citep{PaperVI} and to perform tests on deviations from the Kerr metric~\citep{Psaltis2020,Kocherlakota2021}.

Even though the characteristics of the shadow provide an uncontroversial metric test, fundamentally the observational measurements are based on the properties of the bright image ring, which is used as a proxy for the size of the shadow. Because the image itself is formed through a combination of the spacetime of the black hole and the emission characteristics of the plasma, this inference was justified through an extensive suite of General Relativistic Magnetohydrodynamic (GRMHD) simulations that spanned a wide range of relevant conditions~\citep{PaperV}. These simulations allowed for a quantitative calibration between the diameter of the peak emission, which is measured, and that of the shadow, which is inferred~\citep{PaperVI}.   

There exists a large body of literature exploring the  sensitivity of the horizon scale images, and hence of the metric tests, on the particular assumptions regarding the plasma properties and simulation algorithms employed (see, e.g., \citealt{Dexter2009a,Dexter2010,Dexter2013,Moscibrodzka2009,Moscibrodzka2014,Chan2015,Chan2015b,Medeiros2017,Medeiros2018,Mao2017,Ryan2018,Davelaar2018,Davelaar2019,Narayan2019,Dexter2020,White2020,Yoon2020,Chaterjee2020,Bronzwaer2021,Mizuno2021}), on the possible  deviations from the Kerr metric in the simulated images (see, e.g.,~\citealt{Johannsen2010b,Broderick2014,Johannsen2016,Johannsen2016b,Mizuno2018,Olivares2020}), as well we on the astrophysical complications introduced by the finite resolution of the EHT and the intervening material between the source and the telescopes (see, e.g.,~\citealt{Fish2014,Psaltis2015,Zhu2019}). These studies explored the impact on the resulting images of different magnetic field configurations and initial conditions, of electron heating and acceleration models, of misaligned disk and black hole angular momenta, etc. However, there has not been so far a comprehensive study of the connection between model assumptions, spacetime properties, and image characteristics, especially in the context of metric tests. 

Our goal is to devise a realistic but flexible plasma model in order to explore the impact of the uncertainties in the microphysics and large-scale properties of the accretion flow without relying on the specifics of the GRMHD models. This will also allow us to address recent claims in the literature, made using simple arguments and {\it ad hoc\/} constructions, that ring-like black-hole images could be disjoint from the black-hole shadows (e.g., \citealt{Gralla2019}; \citealt{Gralla2021}). In particular, the uncertainties in the microphysics of plasma heating, the presence of a putative truncation in the accretion flow at an arbitrary radius, the resolution of the simulations, and their initial conditions were all invoked as potential sources of uncertainty that could lead to such images.

In this series of papers, we show that the universal characteristics of the images on which the metric tests are based do not depend on the detailed properties of the numerical simulations, on the plasma model, or on the particular metrics employed but arise naturally in the black hole spacetimes that are characterized by spherical photon orbits and a horizon. In this first paper, we develop a self-consistent analytic model to explore the dependence of image properties on a broad range of plasma characteristics. We demonstrate that, for any flow that obeys basic conservation laws for mass, momentum, and energy, and that is relevant to low-luminosity black holes, the image always closely tracks and straddles the black-hole shadow. Because the black-hole shadow is contained within the width of the ring, when the latter is thin, as is the case in the image observed in M87, we show that the uncertainties in the calibration are small and limited by the fractional width of the observed ring. 

We conclude that the only way to generate an image of the inner accretion flow where the bright emission ring is displaced from the black hole shadow is either as a transient event, such as an Einstein ring~\citep{Chan2015b}, or by artificially truncating the plasma emissivity at an arbitrary radius. Even though the former is a possibility that can be tested by observations that are repeated over many dynamical timescales, the latter violates basic physical considerations. We further show that, for any of the simulated images, the finite resolution of the EHT does not preclude a measurement of the size of the black-hole shadow; it only limits the accuracy of the measurement. In a companion paper~\citep{Younsi2021}, we show that these conclusions do not depend on the Kerr nature of the metric but hold for other metrics as well.

One can of course generate an image structure that is disjoint from the black hole shadow by invoking, e.g., emission from a jet at large distances from the horizon or from the shocks that may be generated in tilted accretion flows \citep{Dexter2013}. However, to be viable, such models will need to account for the stability of the observed images, their thin ring-like structures, and the observed similarity between the diameter of such rings and that of the shadow given the prior mass measurement of the black hole. 

In \S2, we first employ simple emissivity profiles in the accretion flow in order to disentangle the plasma effects from those of the spacetime and identify the conditions necessary to break the coupling between the ring in the image and the black-hole shadow. In \S3, we develop a full analytic plasma model that obeys conservation laws and basic thermodynamic considerations and in \S4, we compare this model to GRMHD simulations. In \S5, we simulate a broad range images based on the analytic plasma model and use them to bound the bias between the diameter of the ring and that of the shadow. 

\begin{figure*} 
 \centerline{
    \includegraphics[width=8.5cm] {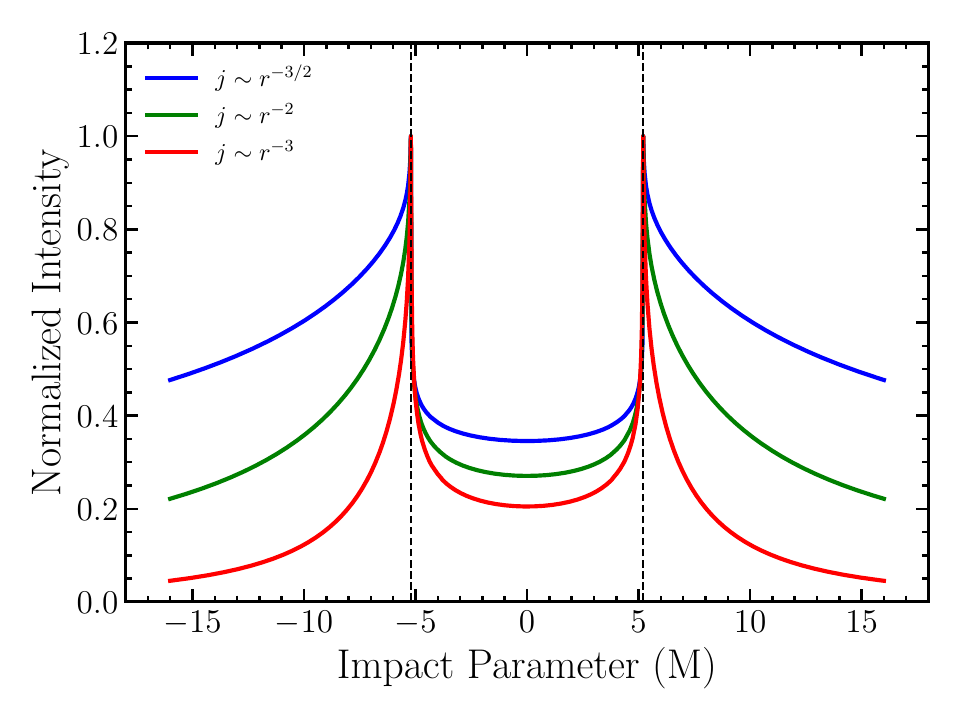}
     \includegraphics[width=8.5cm] {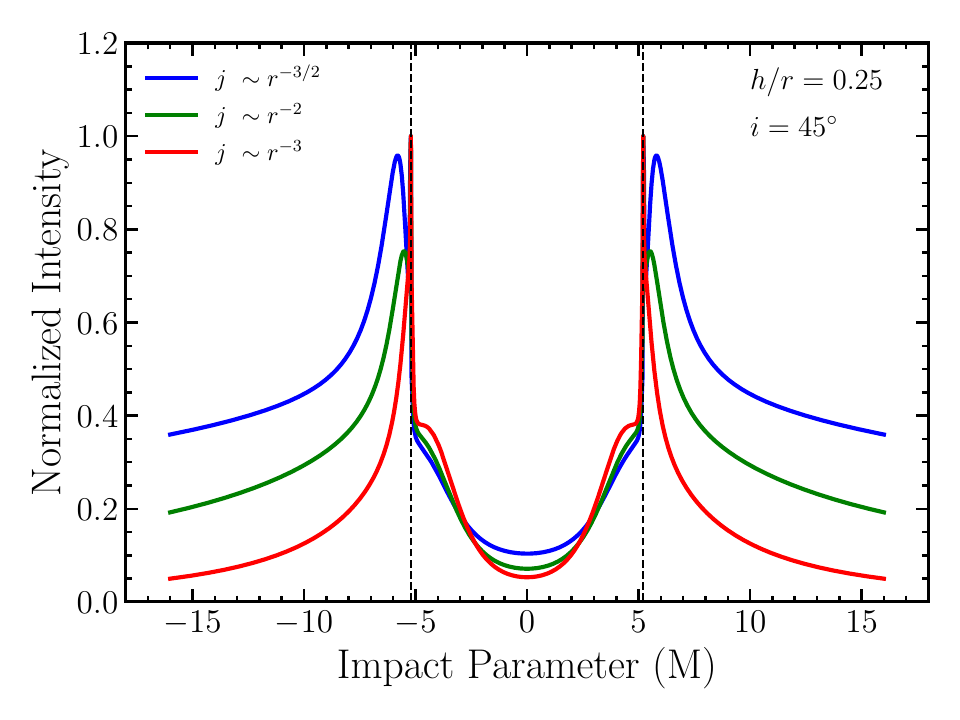}}
 \centerline{
    \includegraphics[width=8.5cm] {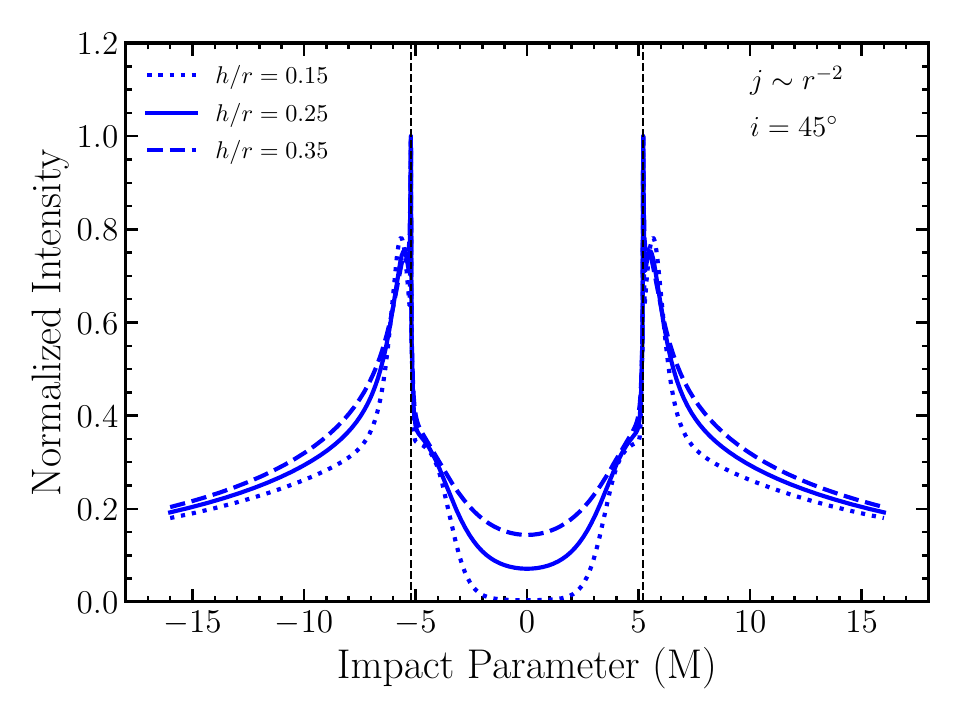}
     \includegraphics[width=8.5cm] {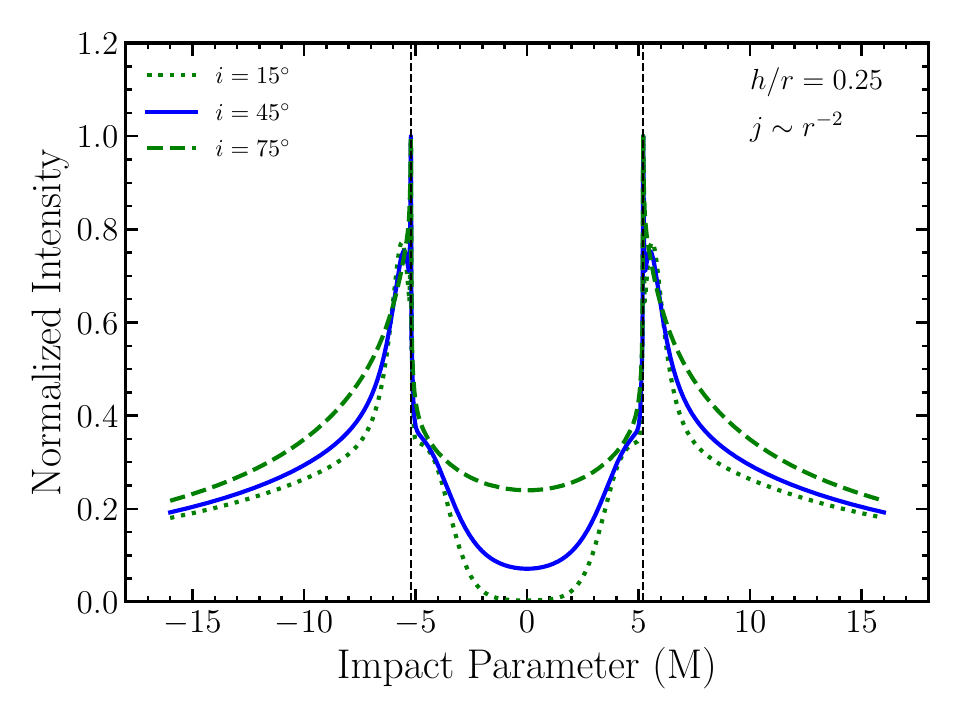}}             
    \caption{\footnotesize Intensity across a horizontal cross section of a black hole image for different analytic emissivity profiles in optically thin accretion flows. 
    {\it(a)} Three radial emissivity profiles for a spherical geometry ($h/r\rightarrow \infty$).  
    {\it(b)} Same as (a) but for a disk scale height $h/r=0.25$.  {\it(c)} Same as (b) but for varying the disk scale height. 
    {\it(d)} Same as (b) but for varying the observer's inclination.
    } 
    \label{fig:analj}
\end{figure*}

\section{General Characteristics of Horizon-Scale Images of Accretion Flows}

Horizon scale images of black holes have a number of universal properties that are shaped by the black hole spacetime and by the physical processes in the plasmas in the accretion flows. In this section, we disentangle the signatures of the spacetime from those of the plasma on the black hole images using a broad range of profiles for the dynamic and thermodynamic properties of the plasma. We focus on the case where the mass of the black hole, the accretion rate, and the observing wavelength are such that the accretion flow is nearly optically thin down to the horizon of the black hole. For the case of the two primary EHT targets, Sgr A* and M87, this occurs at the 1.3 mm wavelength chosen for the EHT observations \citep{Ozel2000}. 

In this paper, we assume that the black hole spacetime is described by the Kerr metric, which in Boyer-Lindquist coordinates is given by 
\begin{eqnarray}
ds^2&=&-\left(1-\frac{2r}{\Sigma}\right)~dt^2
-\left(\frac{4ar\sin^2\theta}{\Sigma}\right)~dtd\phi\nonumber\\
&&\quad +\left(\frac{\Sigma}{\Delta}\right)~dr^2+\Sigma~d\theta^2\nonumber\\
&&\quad+\left(r^2+a^2+\frac{2a^2r\sin^2\theta}{\Sigma}\right)\sin^2\theta~d\phi^2\;.
\label{kerr}
\end{eqnarray}
Here $a$ is the spin of the black hole,
\begin{equation}
\Delta\equiv r^2-2r+a^2,
\end{equation}
and
\begin{equation}
\Sigma\equiv r^2+a^2\cos^2~\theta\;.
\label{deltasigma}
\end{equation}
In this expression, $G=c=M=1$, where $G$, $c$, and $M$ are the gravitational constant, the speed of light, and the black-hole mass, respectively. For the remainder of this section, we will set the spin to zero but will consider the general case in the following sections. In the companion paper, we will consider a variety of non-Kerr metrics.

\begin{figure*} 
  \centerline{
    \includegraphics[width=8.5cm] {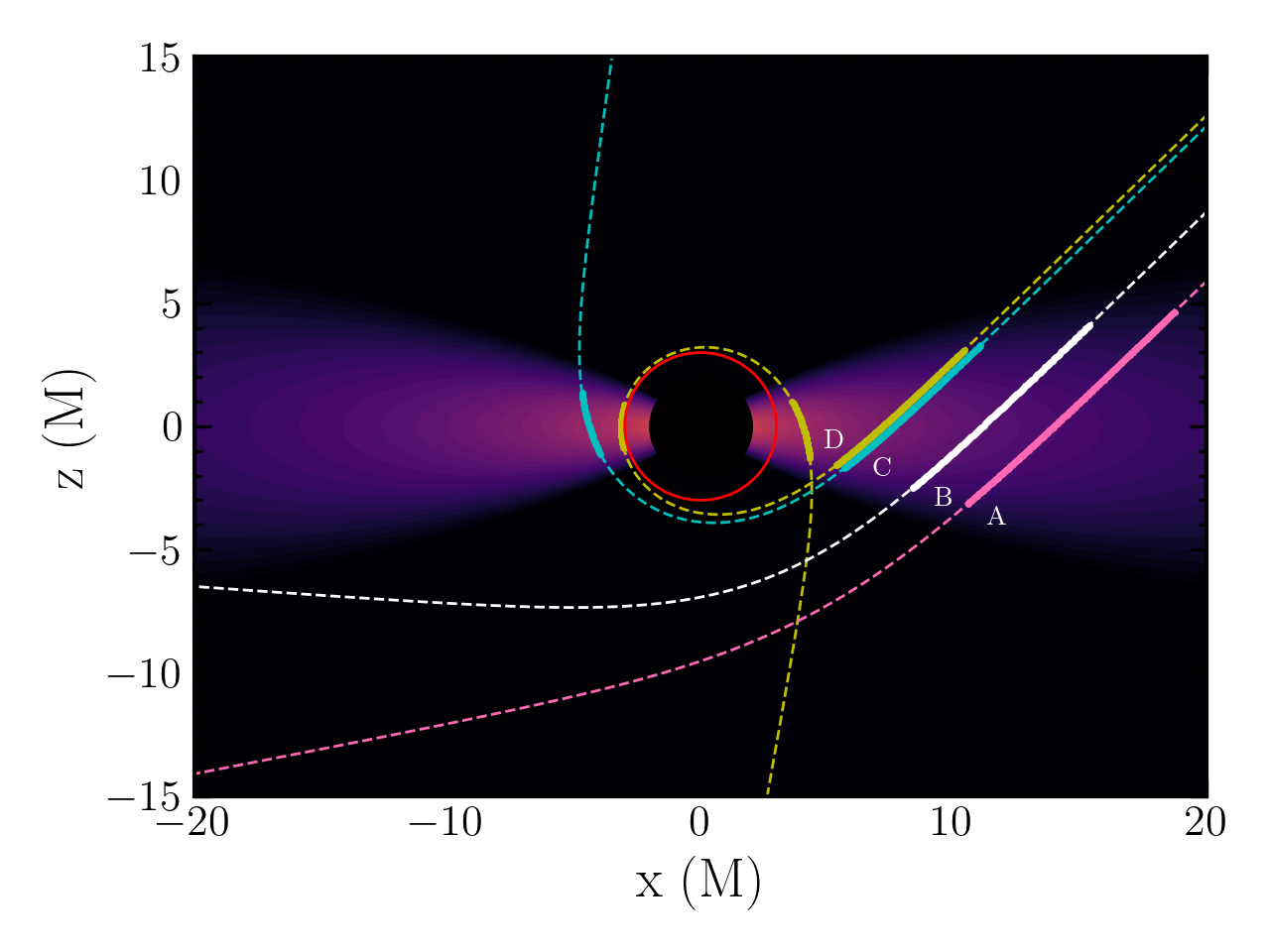}
     \includegraphics[width=8.5cm] {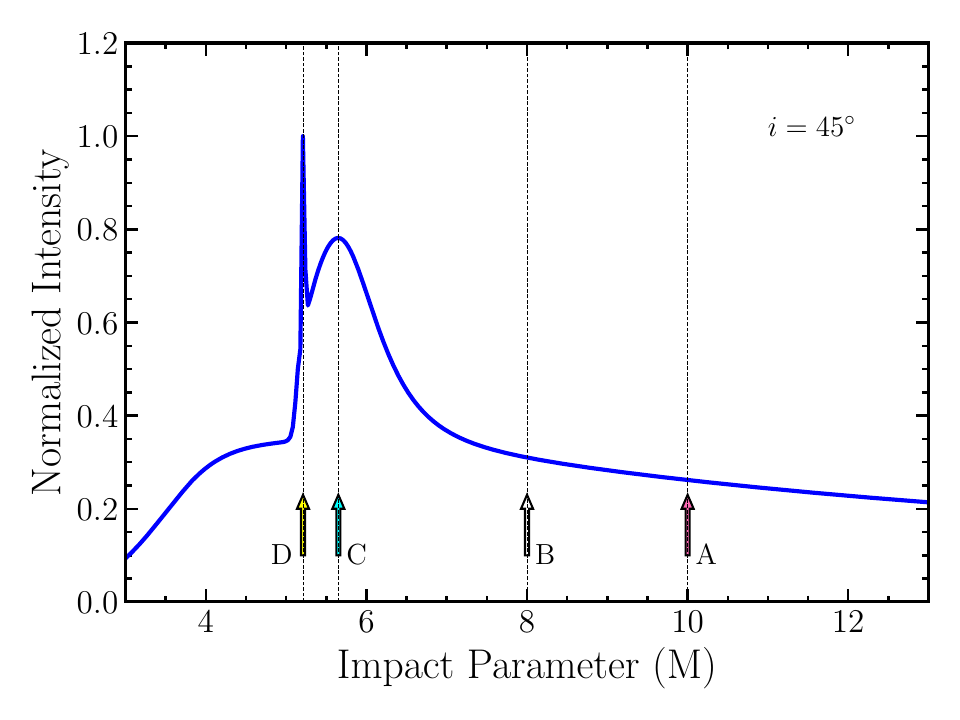}}
    \caption{\footnotesize  {\it (Left)} The different path lengths that contribute to the observed image brightness at four impact parameters through an accretion flow. 
{\it (Right)} The contribution of different impact parameters to the various features in the image cross section.} 
    \label{fig:geodesic}
\end{figure*}

To calculate the images, we place an observer at a large distance and at an inclination $i$ with respect to the spin axis of the black hole. We set an image plane perpendicular to the observer's line of sight and calculate null geodesics backwards from each point on the image plane in the black hole spacetime using the code described in \cite{Psaltis2012}. We then integrate the radiative 
transfer equation along geodesics, which is given by 
\begin{equation}
I(\nu_0) = \int_{\rm ray} j(\nu) g^2 d\lambda
\label{eq:transfer}
\end{equation}
for the case of optically thin emission, where $\lambda$ is the affine parameter, $j_{\nu}$ is the emissivity and $\nu$ is the frequency of the radiation in the local comoving frame, $\nu_0$ is 
the observed frequency, and 
\begin{equation}
g \equiv \frac{\nu_0}{\nu} = \frac{-k_\alpha u^\alpha \vert_\infty}{-k_\alpha u^\alpha}
\label{eq:redshift}
\end{equation}
is the redshift factor~\citep{Dexter2009a}. Here, $k^\alpha$ and $u^\alpha$ are the 4-vectors of the photon momentum and plasma velocity. We have neglected, for simplicity, the effects of self-absorption, which is appropriate for the millimeter wavelength of EHT observations. However, the radio emission at longer wavelengths does become self absorbed and this conceals from view the signature of the black-hole shadow on the image~(see, e.g., \citealt{Narayan1995a,Ozel2000}).

For the purposes of this section, we allow a highly general form of the emissivity that has an arbitrary power-law dependence on the coordinate radius and an arbitrary scale height $h/r$:
\begin{equation}
    j(r,\theta)=j_0 r^{-n}\exp\left[-\frac{1}{2}\left(\frac{\theta-\pi/2}{(h/r)\pi/2}\right)^2\right]\;.
\label{eq:emissivity}    
\end{equation}
When $h/r$ goes to infinity, this describes a spherically symmetric emission geometry. We consider only profiles with $n >1$ because, otherwise, the integral in Eq.~(\ref{eq:transfer}) does not converge and its value is determined entirely by the artificial outer boundary condition. Similarly, for $1 < n \leq 3$, the total flux does not converge, but in this case, the brightness of every pixel is finite on the image plane. In addition, for these toy models, we will mostly consider the flow to be at rest but will also allow for radial infall velocities that are a fraction of the local free-fall velocity. Finally, for the majority of this section, we consider the emissivity to be independent of the photon frequency. In the following section, we will construct self-consistent and physical analytic models of the accretion flow. 

Figure~\ref{fig:analj} shows the cross sections of the brightness of images calculated for a variety of parameters of the simple emissivity model described above. The upper left panel shows the effect of changing the emissivity profile in the flow for a spherically symmetric configuration while the upper right panel makes the same comparison for a geometrically thick disk of $h/r=0.25$ viewed at a 45$^\circ$ inclination. The bottom two panels display the effect of changing the scale height of the disk (left) and the inclination of the observer (right). 

It is evident from these panels that there are universal features of the images that do not depend on the details of the emissivity model:

\noindent {\it (i)} In all cases, there is a precipitous brightness depression interior to the critical impact parameter (i.e., the black hole shadow). This occurs at $\sqrt{27} M$ for non-spinning black holes (see eq.~[\ref{eq:dGR}] for the general expression that shows the marginal effect of spin and observer's inclination). 

\noindent {\it (ii)} The peak brightness always occurs at or very close to this critical impact parameter, again nearly independent of the plasma properties, flow scale height, or the viewing angle.  

\noindent {\it (iii)} Compact images, such as the EHT images of M87, require steep emissivity profiles. 

The main impact of the emissivity profile is to determine the extent of the image at larger impact parameters (i.e., the image compactness). Shallow density profiles lead to extended images that are inconsistent with the compact, narrow ring-like structure observed from the M87 black hole with the EHT. A second aspect of the image that is affected by the emissivity model is the depth of the brightness depression inside the black hole shadow.  However, in none of the cases does this affect the presence or the location of the deep depression. 

\begin{figure*} 
 \centerline{
    \includegraphics[width=8.5cm] {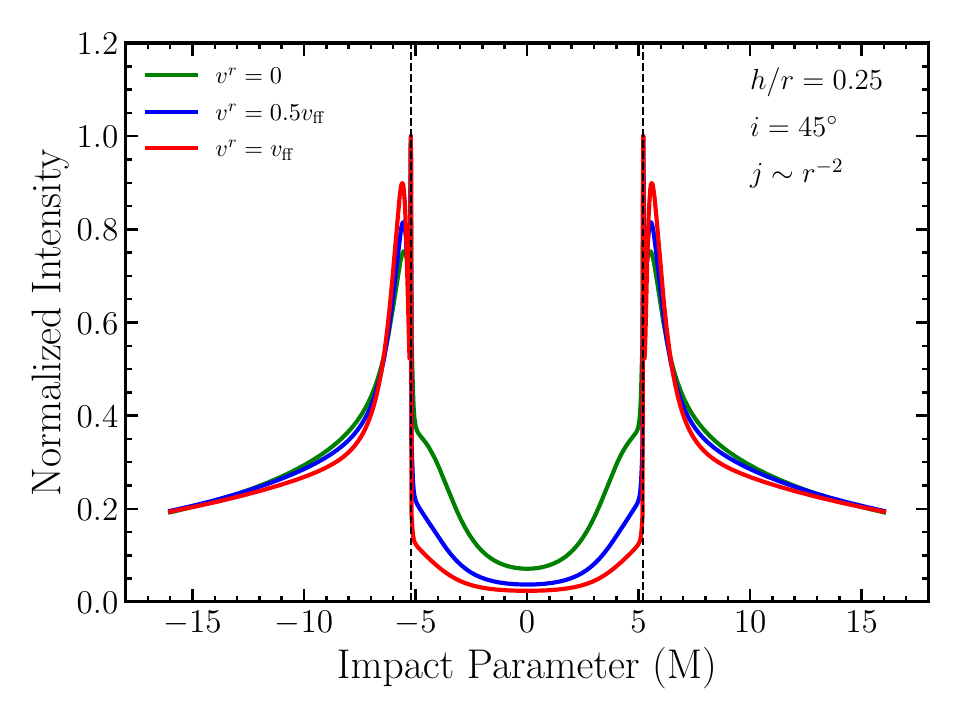}
        \includegraphics[width=8.5cm] {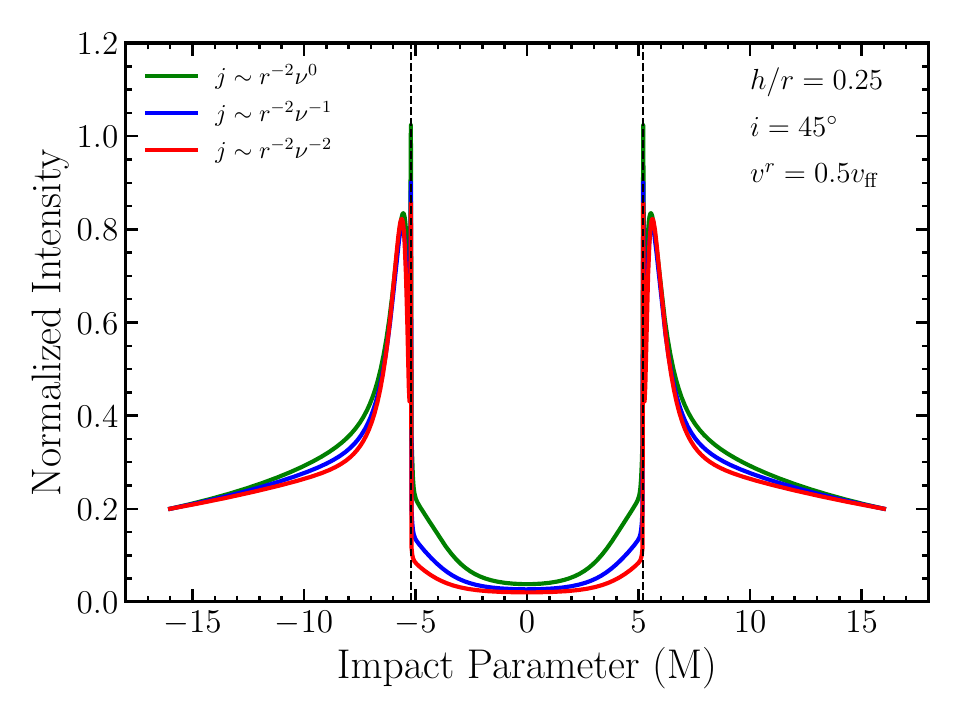}
    }
    \caption{\footnotesize The effect of {\em (Left)\/} a radial infall velocity and {\em (Right)\/} of a frequency-dependent emissivity on the image cross section in the simple analytic model. Both effects cause a reduction of the image brightness at small impact parameters, increasing the contrast between the bright emission ring and the black-hole shadow.} 
    \label{fig:velocity}
\end{figure*}

Figure~\ref{fig:geodesic} elucidates the effect of the emissivity model and the spacetime characteristics on the profile of the image brightness. The brightness seen by an observer at any given impact parameter is determined by the combination of the path length along the ray for which there is non-zero emissivity and the magnitude of that emissivity. This mechanism is different than the image formation in the case of a geometrically thin, optically thick disk (as used in e.g., \citealt{Luminet1979,Gralla2020b,Glampedakis2021}), which is inapplicable to the two primary EHT targets, as thin disks are highly inconsistent with all of their observed characteristics.  

The left panel shows a visualization of the emissivity profile on a meridional plane of the accretion flow and sample trajectories at four impact parameters for an observer placed at $i=45^\circ$. The right panel shows the corresponding image brightness as a function of impact parameter as well as the location of the four trajectories. At large impact parameters, the decline of the image intensity is substantially flatter than the radial dependence of the emissivity profile because of the opposing effects of the decline in emissivity and increase in pathlength with increasing impact parameter.  
In other words, trajectories of A and B give rise to similar intensities because even though the emissivity sampled along path A is smaller than that along path B, the total path length through the densest part of the flow is longer for path A. 

The broad brightness peak around impact parameter C is the result of strong gravitational lensing, which induces a large enough deflection in the photon path to cause it to cross the inner part of the accretion flow twice and, thus, to pick up a larger emissivity contribution. The slight dip between impact parameters C and D is a result of the decreasing disk thickness $h$ with decreasing radius such that, in this example, the lengths of the trajectories, even accounting for the double crossing of the flow, decrease faster inwards than the increase in the emissivity. The situation changes dramatically at a very narrow range of impact parameters around D, for which the trajectories take multiple turns around the photon orbit and give rise to the sharp and narrow brightness peak in the image. Note that the individual number of crossings within the narrow range of critical impact parameter around D will not give rise to distinguishable individual peaks, because the contribution to the radiative transfer integral increases in a continuous manner (and not discretely as in the infinitesimally thin toy models that simply count the number of equatorial crossings). 
 
Figure~\ref{fig:analj} shows that the depth of the central brightness depression (at impact parameters smaller than the critical one) has a dependence on the scale height $h/r$ of the accretion flow and the inclination of the observer. However, the actual brightness at these small impact parameters is heavily overpredicted in this simplistic model because the latter neglects two unavoidable physical effects that further reduce the brightness. First, accretion flows have a finite radial inflow velocity, which becomes a substantial fraction of the speed of light at the small radii intersected by these impact parameters. As a result, the angular dependence of emission is highly peaked towards the black hole, with only a small fraction pointing outward towards the observer. Second, the synchrotron emissivity that gives rise to the radio/millimeter radiation observed from sources such as \sout{M87 and} Sgr~A$^*$ decreases rapidly with increasing photon frequency. The consequence of these redshifts is that the photons that arrive at the observer with a wavelength of 1.3 mm have to be emitted at increasingly higher frequencies when they originate at larger depths in the gravitational potential. The relevant emissivity at smaller radii will, therefore, be smaller than what is assumed in the simple model of Figure~\ref{fig:analj}.

\begin{figure*} 
 \centerline{
    \includegraphics[width=8.5cm] {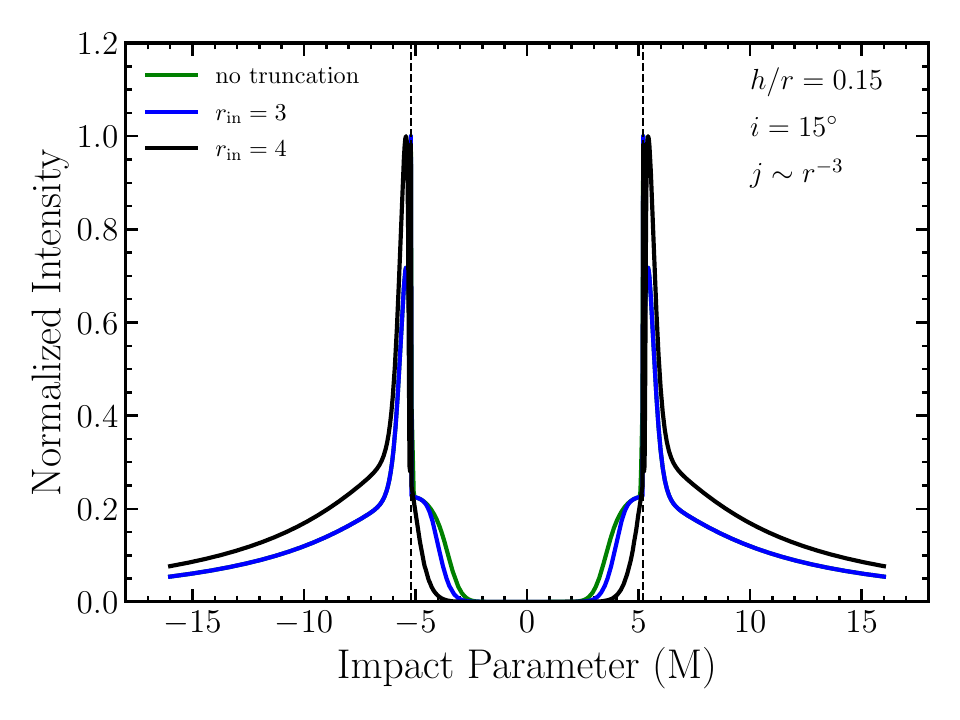}
     \includegraphics[width=8.5cm] {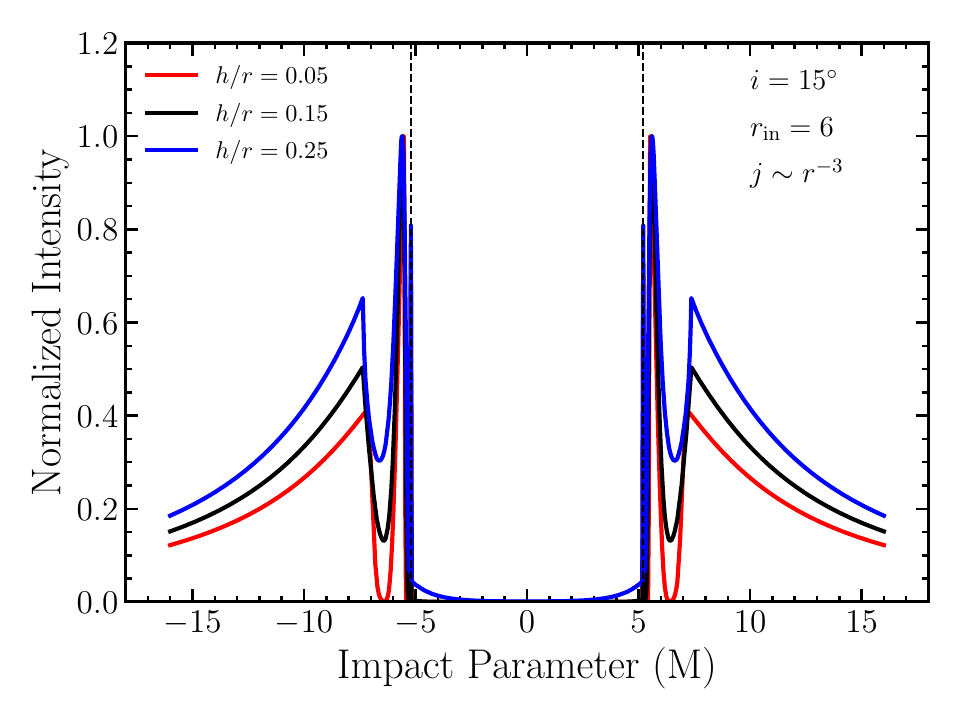}}  
    \caption{\footnotesize Brightness cross sections of images around a non-spinning black hole with the simple emissivity model but for profiles arbitrarily truncated at various radii $r_{\rm in}$. {\it(a)} Images where $r_{\rm in}=3M$ and $r_{\rm in}=4M$ show extremely little difference from those with the horizon boundary condition, even for small values of the disk thickness and observer's inclination, and the sharp drop of image brightness occurs still at the critical impact parameter. {\it(b)} Images with $r_{\rm in} =6M$ exhibit a disjoint feature associated with the truncation radius, in addition to the peak at the photon radius, but the intensity drops to zero between peaks only for extremely small values of the disk thickness. } 
    \label{fig:trunc}
\end{figure*}

The cross sections of the image in the presence of these two effects are shown in Figure~\ref{fig:velocity}. The left panel considers a radial inflow velocity that is a fraction of the local free-fall velocity ($u^r_{\rm ff}=\sqrt{2/r}$), while the right panel introduces a modest power-law dependence of the synchrotron emissivity (note that at high frequencies, the frequency dependence of the synchrotron emissivity is in fact exponential). As expected, in both cases, the addition of these necessary physical effects substantially reduces the brightness inside the black hole shadow and further enhances the sharp drop at the critical impact parameter. 

These simple models demonstrate that for any continuous plasma distribution in the accretion flow, compact images of black holes are always characterized by a large brightness depression at the critical impact parameter independent of any details of the plasma or its emission. This brightness depression is a unique signature of the spacetime that cannot be overwhelmed by plasma effects for the conditions of the primary EHT targets. Further, its diameter is determined entirely by the size of the photon orbit and the strength of gravitational lensing. 

Given the generality of the conclusions above, one is left to ask what conditions would be required to generate a compact ring that is disjoint from the location of the shadow. The only remaining mechanism for decoupling the brightness depression from the critical impact parameter is the introduction of an ad hoc truncation of the emissivity profile at some arbitrary radius. We explore the consequences and the feasibility of such truncations in the remainder of this section and in \S5. 

\subsection{The effect of an arbitrary truncation of disk emissivity}

We turn to the images associated with an accretion flow where the emissivity profile is truncated at an arbitrary inner boundary that is separate from the horizon of the black hole. Previous studies (e.g., \citealt{Luminet1979,Straub2012,Vincent2015}) have invoked such ad hoc conditions by setting the matter density to zero at, e.g., the innermost stable circular orbit (ISCO), motivated by the fact that matter loses centrifugal support for a cold, thin disk at that radius. In principle, such a configuration can also be produced by setting the electron temperature or the magnetic field strength suddenly to negligible values at a chosen radius. 

Figure~\ref{fig:trunc} shows the image cross sections obtained from the simple analytic model truncated at a variety of inner disk radii $r_{\rm in}$ and for different values of the disk thickness. In all cases, the emissivity profile is given by $j \sim r^{-3}$, the observer's inclination is set to 15$^\circ$, and the radial velocity is $v_r = 0.5 v_{\rm ff}$. The left panel shows that, if the truncation is close to the radius of the photon orbit ($r_{\rm in} \lesssim 5\;M$), then its radius has minimal effects on the shadow characteristics; the sharp drop in the image brightness still occurs at the critical impact parameter. The only way to generate a disjoint peak of emission at a substantially different impact parameter is by truncating the emissivity at a much larger radius. Note that, if such a truncation is associated with the ISCO, it will be significantly displaced from the photon ring only for slowly spinning black holes, as the coordinate radius of the ISCO and that of the photon orbit converge to the same value as the spin increases to its maximum value. 

The right panel of Figure~\ref{fig:trunc} shows a case with $r_{\rm in} = 6M$ for three different values of the disk thickness. In this case, multiple disjoint peaks are produced. The first peak still appears near the critical impact parameter, while the outer peak associated with the truncation of the disk has a location that depends on the disk thickness $h/r$.\footnote{The claim in \citet{Gralla2020a} that this peak appears at $r_{\rm in}+1$ is correct only in the highly specialized case of face-on observers and purely equatorial emission.} Even with this large truncation radius, the depth between the individual peaks of emission is also a function of the disk thickness (and to a lesser extent, of the observer's inclination and the radial dependence of the emissivity, not shown in the figure).  

In the next section, we develop a self-consistent, semi-analytic, covariant model of the accretion flow in order to explore the physicality of such truncated emissivity profiles. 

\section{The Properties of the Plasma in the Accretion Flow}

A realistic analytic model for the plasma that is needed to calculate horizon-scale images of the accretion flow requires the specification of the fluid density $\rho$, the four-velocity $u^\mu$, magnetic field strength $B$, and the electron temperature $T_{\rm e}$. These, in turn, are obtained from the more general basic conservation equations for the stress-energy tensor $T^{\mu \nu}$, which are 
\begin{equation}
T^{\mu \nu}_{;\nu} = 0
\label{eq:consE}
\end{equation}
for the conservation of energy-momentum and 
\begin{equation}
(\rho u^\mu)_{;\nu} = 0
\label{eq:consM}
\end{equation}
for the conservation of particle number or, equivalently, rest-mass density. These conservation laws are satisfied independent of whether the plasma is in the kinetic or the fluid regime, of the validity of the MHD approximation, or of the particular dissipation mechanisms involved. 

By construction, our analytic model is axisymmetric so that there is no dependence of any quantity on the azimuthal angle $\phi$. Furthermore, in this section, we use vertically-averaged quantities for the accretion flow such that the conservation equation for the mass density becomes 
\begin{equation}
4 \pi \left(\frac{h}{r}\right)\sqrt{-g}\rho u^r = - \dot{M}, 
\label{eq:cont_gen}
\end{equation}
where $\dot{M}$ is the constant mass accretion rate throughout the flow and the scale height is defined as
\begin{equation}
    \frac{h}{r}(r)\equiv \frac{\int_0^{2\pi} d\phi \int_0^\theta d\theta \vert \pi/2-\theta\vert \rho \sqrt{-g}}{\int_0^{2\pi} d\phi \int_0^\theta d\theta \; \rho \sqrt{-g}}\;.
\end{equation}
In eq.~[\ref{eq:cont_gen}], the factor related to the determinant of the metric, $g=\vert \det{g_{\mu\nu}}\vert$, is understood to be evaluated at the equatorial plane. For the Kerr metric, $\sqrt{-g}=r^2$ on the equatorial plane and, therefore, the above equation becomes
\begin{equation}
4 \pi \left(\frac{h}{r}\right)r^2\rho u^r = - \dot{M}. 
\label{eq:cont}
\end{equation}
The electron density as a function of radius is then simply 
\begin{equation}
n_{\rm e}(r) = \frac{\rho(r)}{m_{\rm p}} = \frac{\dot{M}}{4 \pi r^2 (h/r) u^r m_{\rm p}}, 
\label{eq:ne_gen}
\end{equation}
where $m_{\rm p}$ is the mass of the proton (assuming a fully ionized hydrogen plasma) and $u^r$ is the $r-$component of the plasma velocity. It is evident from this equation that the density profile only depends on the radial component of the four-velocity, which we will specify below. 

In a radiatively inefficient flow, energy conservation implies that the energy content of the fluid is determined by viscous and compressional heating. Following \citet{Gammie98}, we write the stress-energy tensor of matter as 
\begin{equation}
T^{\mu \nu} = P g^{\mu \nu} + (\rho + \epsilon +P) u^\mu u^\nu + t^{\mu \nu},
\label{eq:Tmunu}
\end{equation}
where $\epsilon$ and $P$ are the internal energy and pressure of the plasma, respectively, and $t^{\mu \nu}$ is the  stress-energy tensor associated with the dissipation mechanism. 

We consider three contributions to the pressure and internal energy: the ions, the electrons, and the stress-energy tensor of the electromagnetic field present in an MHD flow. We write the total pressure $P$ as the sum of gas and magnetic pressures
\begin{equation}
P = P_{\rm g} + P_B = n_{\rm i} kT_{\rm i} + n_{\rm e} k T_{\rm e} + P_B\;,
\end{equation}
where $n_{\rm i}$ and $n_{\rm e}$ are the ion and electron number densities, respectively, and $T_{\rm i}$ and $T_{\rm e}$ are their temperatures. By adopting this form of the pressure, we make the assumption that the velocity distribution of the particles in the comoving frame is predominantly Maxwellian. Defining the ratio of the ion-to-electron temperature as $R$ and the plasma-$\beta$ as $\beta=P_g/P_B$, we write the total pressure as 
\begin{equation}
P=n_{\rm i} k T_{\rm i} \frac{R+1}{R} \frac{\beta+1}{\beta}. 
\end{equation}
For the internal energy, we adopt the following form 
\begin{equation}
\epsilon =  \frac{1}{\hat{\gamma}-1} \frac{\rho k_{\rm B} T_{\rm i}}{m_{\rm p}} \frac{R+1}{R}.
\label{eq:epsilon}
\end{equation}
For a purely ionized hydrogen plasma, this equation describes the internal energy, when $\hat{\gamma}=5/3$. When we incorporate the contribution due to the magnetic field, we keep the same expression but allow for $\hat{\gamma}$ to be the effective adiabatic index of the magnetized plasma. 

The equation for energy conservation can be obtained by projecting equation~(\ref{eq:consE}) onto the four-velocity of the plasma 
\begin{equation}
u^\mu T^{\mu \nu}_{;\nu} = 0.
\label{eq:cons3} 
\end{equation}
For the form of the stress-energy tensor introduced above, this reduces to 
\begin{equation}
u^r \frac{d\epsilon}{dr} - u^r \frac{\epsilon+P}{\rho}\frac{d\rho}{dr} = \Phi - \Lambda, 
\label{eq:Econs}
\end{equation}

For a radiatively inefficient flow, $\Lambda$ is negligible by definition. This assumption has been verified in a number of GRMHD simulations in which the effects of radiative cooling have been explicitly considered (see, e.g., \citealt{Ryan2018}). In simulations with even the relatively high mass accretion rates inferred for the M87 black hole, the effect of radiative cooling leads to a reduction of ion temperature near the event horizon by only 10\% and a reduction of the scaleheight of the flow by about 5\%. These small changes are within the uncertainties of our semianalytic model and do not affect qualitatively our results.

Inserting the expressions for the pressure and internal energy and performing some algebraic manipulations, we cast this equation~(\ref{eq:Econs}) in the form
\begin{equation}
\frac{d}{dr} \left[T_{\rm i} \rho^{-\frac{(\beta+1)(\hat{\gamma}-1)}{\beta}}\right] = \frac{m_{\rm p} R (\hat{\gamma}-1)}{k_{\rm B} (R+1)} \frac{\Phi}{u^r \rho} \rho^{-\frac{(\beta+1)(\hat{\gamma}-1)}{\beta}},
\label{eq:Ti_dis}
\end{equation}
where $k_{\rm B}$ is the Boltzmann constant. Using the continuity equation~(\ref{eq:cont_gen}) to write the product $\rho u^r$ in terms of the mass accretion rate and integrating it from infinity down to radius $r$, we obtain for the ion temperature  
\begin{equation}
T_{\rm i} = \frac{m_p R (\hat{\gamma}-1)}{k_{\rm B} (R+1)} {\cal V}, 
\label{eq:Ti_int}
\end{equation}
where we have defined the quantity
\begin{equation}
{\cal V} \equiv \rho^{\frac{(\beta+1)(\hat{\gamma}-1)}{\beta}} \int_\infty^r \frac{\Phi}{\dot{M}} \rho^{-\frac{(\beta+1)(\hat{\gamma}-1)}{\beta}} \left(\frac{h}{r}\right) 4 \pi \sqrt{-g} dr. 
\label{eq:calV}
\end{equation}
This quantity is a volume integral throughout the accretion flow of the dissipation rate per particle, weighted by a power-law function of the density. If this weighting function was equal to unity, the integral would simply be comparable to the potential energy of a test particle at radius $r$. 

In Appendix~A, we calculate the volume integral in equation~(\ref{eq:calV}) for specific models of energy dissipation in the accretion flow by viscous stresses. As we show there, the radial profile of the density, the black-hole spin, and other parameters introduce only subdominant effects. Indeed, given that all the quantities in the integral have practically power-law dependences on radius, the effect of the weighting function is to introduce a multiplicative factor that is nearly constant and of order unity. As a result, we can write the ion temperature for the Kerr metric in the form
\begin{equation}
    T_{\rm i}= \frac{m_p c^2}{k_{\rm B}} \frac{R (\hat{\gamma}-1)}{(R+1)} \zeta \left(\frac{GM}{rc^2}\right), 
\label{eq:Ti}
\end{equation}
where $\zeta$ is an order-unity factor. Following our definitions above, we write the electron temperature as
\begin{equation}
    T_{\rm e}=\frac{m_p c^2}{k_{\rm B}} \frac{(\hat{\gamma}-1)}{(R+1)} \zeta \left(\frac{GM}{rc^2}\right)\;.
\label{eq:Te}
 \end{equation}
We also obtain the magnitude of the magnetic field from the definition of plasma-$\beta$ as
\begin{equation}
B= \left[\frac{8\pi}{\beta} \frac{\rho}{(\hat{\gamma}-1) m_{\rm p}} \frac{R+1}{R}k_{\rm B}T_i\right]^{1/2}\;,
\end{equation}
which becomes
\begin{equation}
B= \left[\frac{8\pi}{\beta}\zeta \left(\frac{GM}{rc^2}\right)\rho c^2\right]^{1/2}\;.
\end{equation}

Using the density, temperature, and the magnetic field strength calculated above, we can now evaluate the synchrotron emissivity from a thermal and isotropic distribution of relativistic electrons. We use the analytic fitting formula for the angle-averaged emissivity derived by~\citet{Mahadevan1996}, which is accurate to within 2.6\% for all temperatures and frequencies of interest:
\begin{equation}
 \label{eq:jtot}
 j_\nu = \frac{n_{\rm{e}} e^2}{\sqrt{3} c K_2(1/\theta_e)} \nu M(x_M), 
\end{equation}
with $M(x_M)$ given by
\begin{eqnarray}
M(x_{M}) = \frac{4.0505~\mathpzc{a}}{x_{M}^{1/6}} \left(1 + \frac{0.40~\mathpzc{b}}{x_{M}^{1/4}} +  
     \frac{0.5316~\mathpzc{c}}{x_{M}^{1/2}}\right) \\ \nonumber
     \times \exp(-1.8896 \; x_{M}^{1/3}). 
 \label{eq:mxm}     
\end{eqnarray}
Here, $\nu_b \equiv e B / 2 \pi m_e c$ is the cyclotron frequency,
\begin{equation}
 \label{eq:xm}
 x_M \equiv \frac{2 \nu}{3 \nu_b \theta_e^2}, 
\end{equation}
$\theta_{\rm e} \equiv k T_{\rm e} / m_{\rm e} c^2$ is the dimensionless electron temperature, and $K_2(x)$ is the modified Bessel function of the second kind. The best fit values of the coefficients $\mathpzc{a}, \mathpzc{b}$, and $\mathpzc{c}$ for different temperatures are given in \citet{Mahadevan1996}. 

Finally, we calculate the disk scale height. Under the assumption of hydrostatic equilibrium in the vertical direction, \citet{Abramowicz96} and \citet{Gammie98} derive
\begin{equation}
\left(\frac{h}{r}\right)^2 = \frac{P}{(\rho+\epsilon+P) r^2 \nu_z^2},
\end{equation}
where $\nu_z$ is an effective frequency of vertical oscillations. For a relativistic thin disk in a Kerr spacetime, this reduces to 
\begin{equation}
\left(\frac{h}{r}\right)^2 = \frac{P r}{(\rho+\epsilon+P)} \left[\frac{1-3r^{-1}+2ar^{-3/2}}{1-4a r^{-3/2}+3a^2r^{-2}}\right]
\end{equation}
The last term describes the relativistic corrections, which become significant only near the location of the photon orbit, where the assumption of hydrostatic equilibrium clearly breaks down. On the other hand, away from the photon orbit and under the assumptions of our model, we can write
\begin{equation}
\frac{h}{r} = \frac{1}{r u^\phi} \left(\frac{P}{\rho}\right)^{1/2} = \sqrt{(\hat{\gamma}-1)\zeta}. 
\label{eq:hr}
\end{equation}
As we showed in the previous section, the scale height affects only the details of the image structure and not the presence or the location of the brightness depression. For this reason, we proceed with this last expression that depends only on constants, unless we specify otherwise. 

\subsection{Plasma Velocities}

The use of the conservation laws in the previous section demonstrated that the radial profiles of the various plasma quantities are determined almost entirely by the radial velocity profile in the radiatively inefficient accretion flows. Simulations and numerous analytic arguments have shown that outside the radius of the innermost stable circular orbit $r_{\rm ISCO}$, the plasma orbits with velocities comparable to the test particle angular velocities and drifts inwards because of the outward transport of angular momentum. Inside the ISCO, circular orbits are unstable and the plasma plunges towards the horizon. We now describe our model for the velocity profiles that accounts for this qualitative behavior. 

Throughout the flow, we set the $\theta-$component of the velocity equal to zero, i.e., we assume that the plasma has orbital and radial drift velocities only. In addition, because of the assumed axisymmetry, the velocity vector will depend only on the spherical coordinates $(r,\theta)$.

First, we calculate the velocity profile in the equatorial plane, i.e., at $(r,\theta=\pi/2)$. 

\noindent {\it Outside the ISCO radius.---\/} We  assume that the azimuthal velocity of plasma is equal to the local Keplerian orbital velocity. To calculate the latter, for a general axisymmetric metric written in Boyer-Lindquist-like coordinates, we first write the general expression for the angular velocity of a test particle, as measured by an observer at infinity~\citep{Ryan1995}
\begin{equation}
\Omega(r)=\frac{-g_{t\phi,r}+\sqrt{(g_{t\phi,r})^2-g_{tt,r}g_{\phi\phi,r}}}{g_{\phi\phi,r}}\;,
\label{eq:omega}
\end{equation}
and use this to calculate the equatorial azimuthal velocity outside the ISCO as
\begin{equation}
    u^\phi_{eq}(r)=\frac{\Omega}{\sqrt{-g_{tt}-(2g_{t\phi}+g_{\phi\phi}\Omega)\Omega}}, 
\end{equation}
where commas denote ordinary differentiation, as usual. 

The radial component of the plasma velocity depends on the Reynolds and Maxwell stresses, which control the rate of outward angular momentum transport, as well as on the plasma pressure and magnetic stresses that may provide support against plunging towards the black hole. To allow for a general form of the radial velocity profile that does not depend on the specifics of angular momentum transport, we write 
\begin{equation}
    u^r_{eq}(r)=-\eta \left(\frac{r}{r_{\rm ISCO}}\right)^{-n_r}
    \label{eq:vr}
\end{equation}
where $\eta$ and $n_r$ are free parameters. Note that when $n_r=1/2$, the radial velocity becomes a fraction $\eta$ of the azimuthal velocity at large radii (i.e., the Newtonian profile). 

Finally, we compute the $t-$component of the velocity by imposing the requirement 
\begin{equation}
\vec{u}\cdot\vec{u}=g_{tt} (u^t)^2+g_{rr} (u^r)^2+g_{\phi\phi} (u^\phi)^2+2g_{t\phi} u^t u^\phi=-1\;.
\label{eq:vt}
\end{equation}

\begin{figure*} 
 \centerline{
    \includegraphics[width=\textwidth/3] {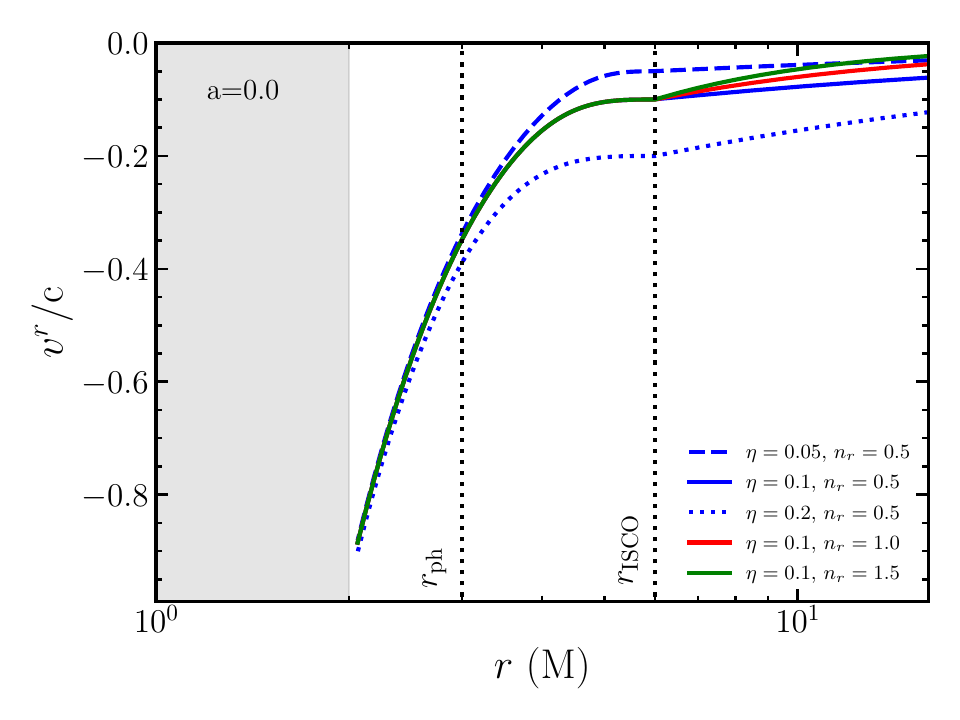}
     \includegraphics[width=\textwidth/3] {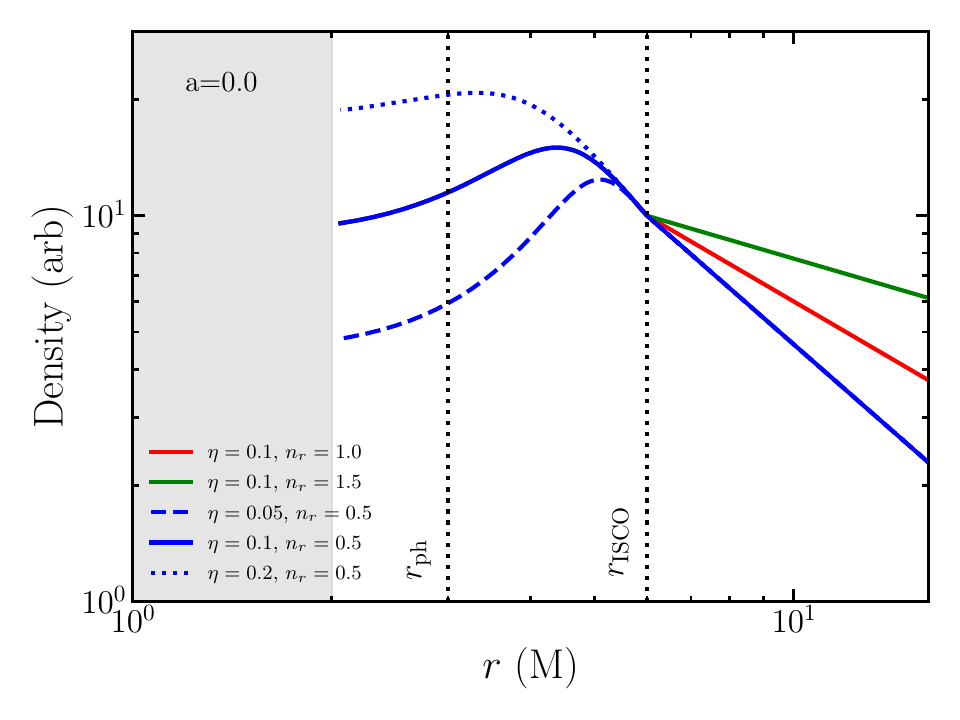}
     \includegraphics[width=\textwidth/3] {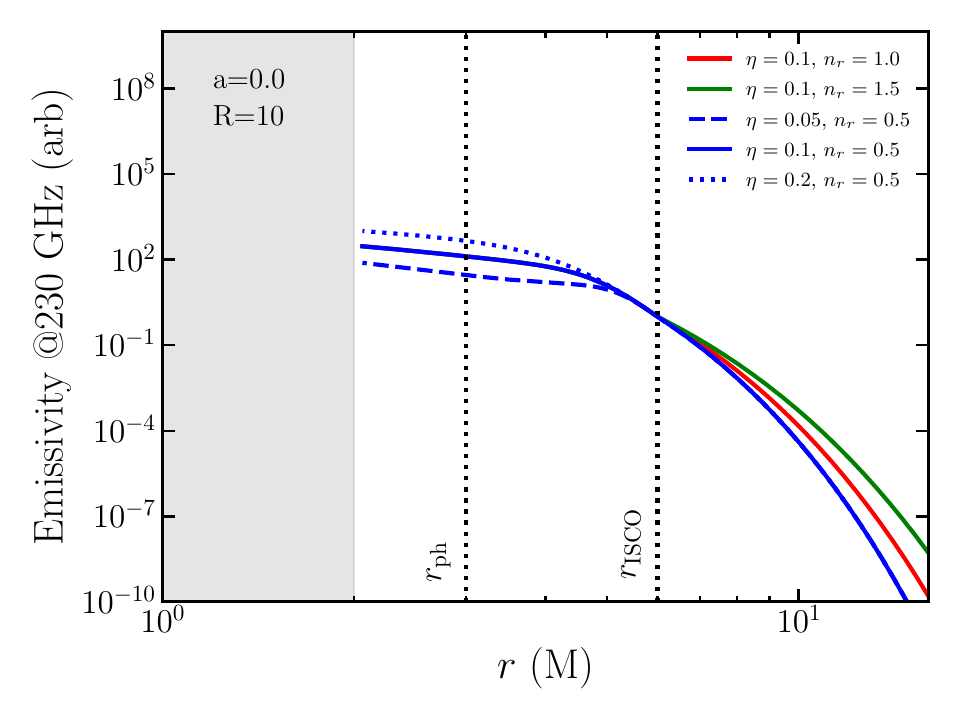}}
 \centerline{
    \includegraphics[width=\textwidth/3] {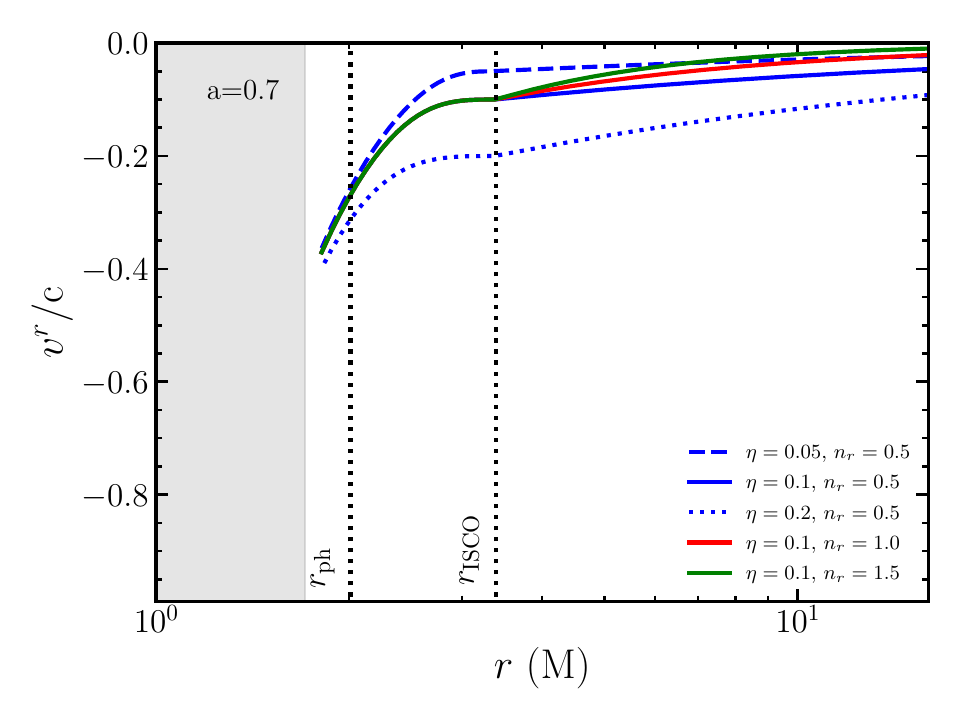}
     \includegraphics[width=\textwidth/3] {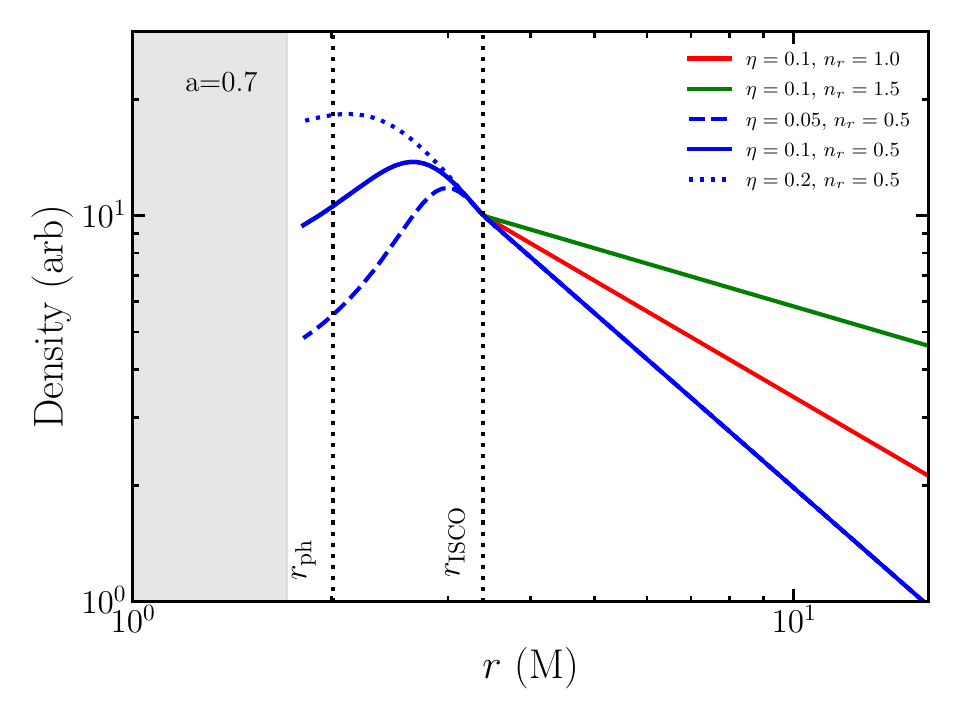}
     \includegraphics[width=\textwidth/3] {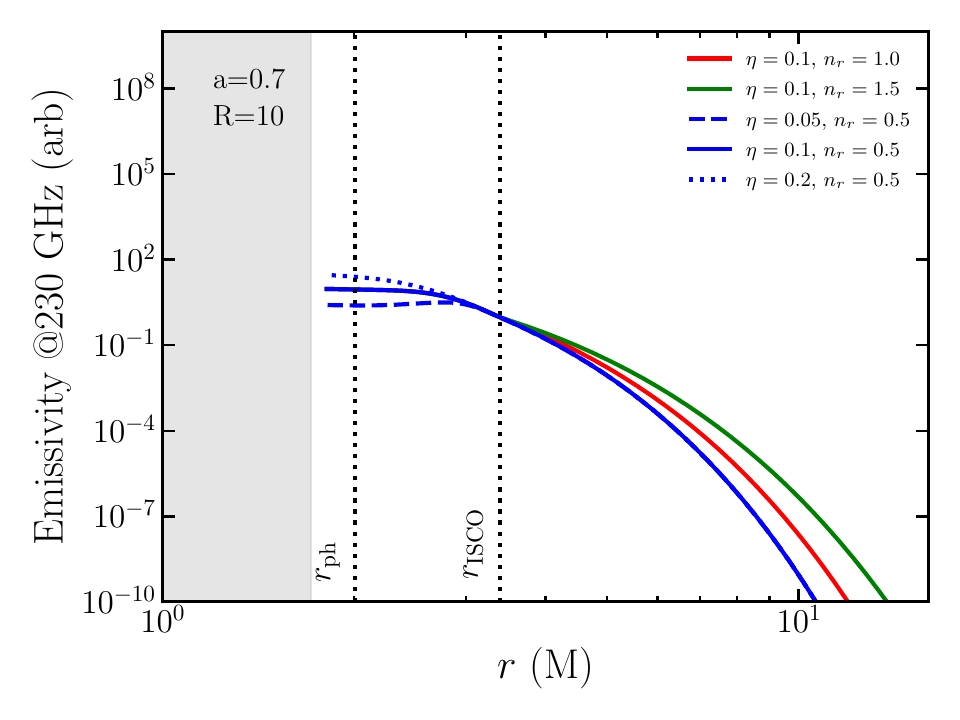}}
    \caption{\footnotesize The {\it (Left)\/} equatorial radial velocity, {\it (Middle)\/} density, and {\it (Right)\/} the comoving 230~GHz synchrotron emissivity profiles of accretion flows around Kerr black holes with spins equal to {\it (Top)\/} $a=0$ and {\it (Bottom)\/} $a=0.7$. In all panels, the region inside the horizon is shown with a gray-filled area and the coordinate radii of the photon obit ($r_{\rm ph}$) and of the ISCO ($r_{\rm ISCO})$ are shown with vertical dashed lines. Curves of different colors correspond to different radial profiles outside the ISCO whereas curves of different linestyles correspond to different magnitudes of the radial velocity at the ISCO. In the rightmost panels, the ion-to-electron temperature ratio is set to $R=10$ and the magnetic field strength at the ISCO to $B_{\rm ISCO}=10$~G. The density and emissivity have been normalized to their respective values at the ISCO. Even though the magnitude of the radial velocity increases rapidly inside the ISCO, conservation of mass and energy throughout the flow causes the 230~GHz emissivity to always increase inwards. There is no evidence for a sharp reduction in emissivity with decreasing radius, inside the ISCO or anywhere else in the flow.} 
    \label{fig:profiles}
\end{figure*}

\noindent {\it At the ISCO radius.---\/} We calculate the energy and angular momentum of the plasma (and not of the test particles) at the ISCO, i.e., accounting for the radial velocity, using
\begin{equation}
    E_{{\rm ISCO}}=-g_{tt} u^t_{\rm eq}(r_{\rm ISCO}) - g_{t\phi}u^\phi_{\rm eq}(r_{\rm ISCO})
\end{equation}
and
\begin{equation}
    L_{\rm ISCO}=g_{t\phi} u^t_{\rm eq}(r_{\rm ISCO}) + g_{\phi\phi}u^\phi_{\rm eq}(r_{\rm ISCO}),
\end{equation}
where the right-hand sides of these equations have been calculated at the equatorial plane.

\noindent {\it Inside the ISCO radius.---\/} Inside the ISCO, the plasma loses centrifugal support and plunges towards the horizon. In the absence of any material or magnetic stresses, the plunging occurs along the geodesics of the spacetime and is simply described by the velocities of the free-falling test particles, with the energy and the angular momentum of the plasma evaluated at the ISCO radius. On the equatorial plane, this gives for the azimuthal component of the velocity 
\begin{equation}
    u^\phi_{eq}(r)=\frac{g_{tt} L_{\rm ISCO}+g_{t\phi} E_{\rm  ISCO}}{g_{tt}g_{\phi\phi}-g_{t\phi}^2}
\end{equation}
and 
\begin{equation}
u^t_{eq}(r)=-\frac{g_{t\phi} L_{\rm ISCO}+g_{\phi\phi} E_{\rm ISCO}}{g_{tt}g_{\phi\phi}-g_{t\phi}^2}    
\end{equation}
for the $t-$component. Finally, we calculate $u^r_{\rm eq}(r)$ from the condition $\vec{u}\cdot \vec{u}=-1$, 
as before. 

\subsection{Can Plunging Inside the ISCO Truncate the Emissivity?}

In Figure~\ref{fig:profiles}, we show the radial profiles of the radial velocity $u^r$, density $\rho$, and the resulting 230~GHz synchrotron emissivity $j$ for the covariant analytic models described above, for two values of the black hole spin parameter $a$. In each panel, radial profiles are shown for the flow outside the ISCO for three values of the parameter $n_r$ in eq.~(\ref{eq:vr}) and plunging solutions are shown interior to $r_{\rm ISCO}$ for different normalization of the radial velocity at the ISCO. Even though the magnitude of the radial velocity increases rapidly inside the ISCO because the plasma loses centrifugal support (left panels), conservation of mass in the converging flow prevents the density from decreasing significantly (or at all) in the plunging region (middle panels). Furthermore, the conservation of energy and magnetic flux causes the electron temperature and the magnetic field strength to increase rapidly inside the ISCO such that, in all cases, the 230~GHz emissivity continues to increase inwards. We also note that this characteristic rapid inward rise of the synchrotron emissivity in inner accretion flows, in conjunction with the rapidly increasing photon pathlengths toward the photon ring, plays an important role in the formation of the thin rings in horizon-scale images of black holes, as discussed in \S2. 

In the previous section, we showed that a sudden truncation of the emissivity is required to generate a brightness depression in the image that is disjoint from the black hole shadow. We conclude here that, in self-consistent analytic models that obey conservation laws, the loss of centrifugal support and the rapid plunging of the plasma inside the ISCO do not produce such a truncation in the emissivity. 

\begin{figure*} 
 \centerline{
    \includegraphics[width=8.5cm] {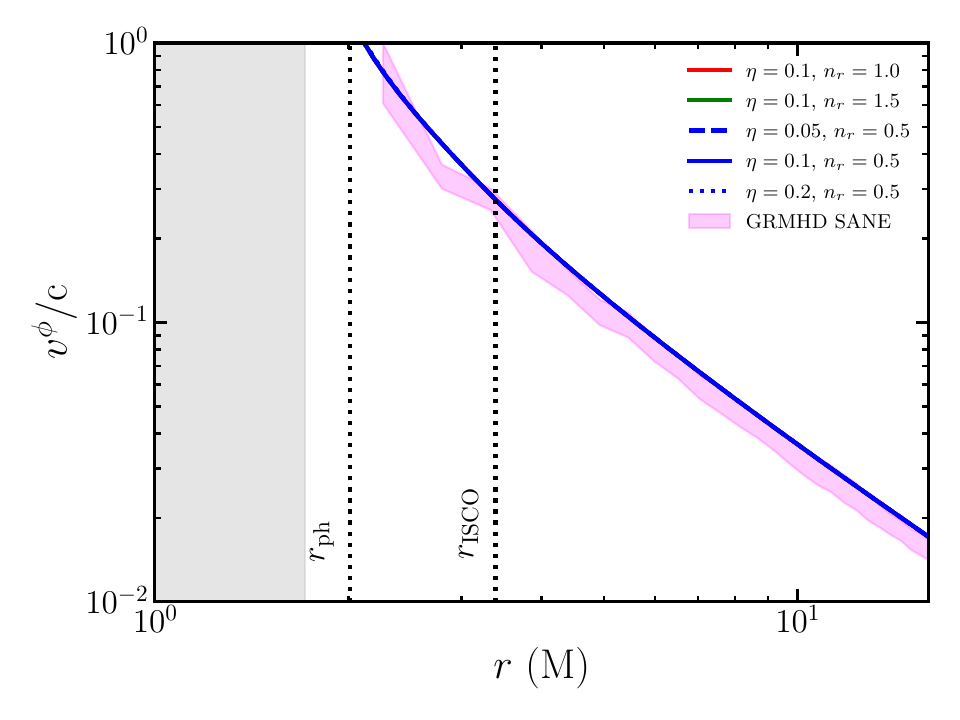}
     \includegraphics[width=8.5cm] {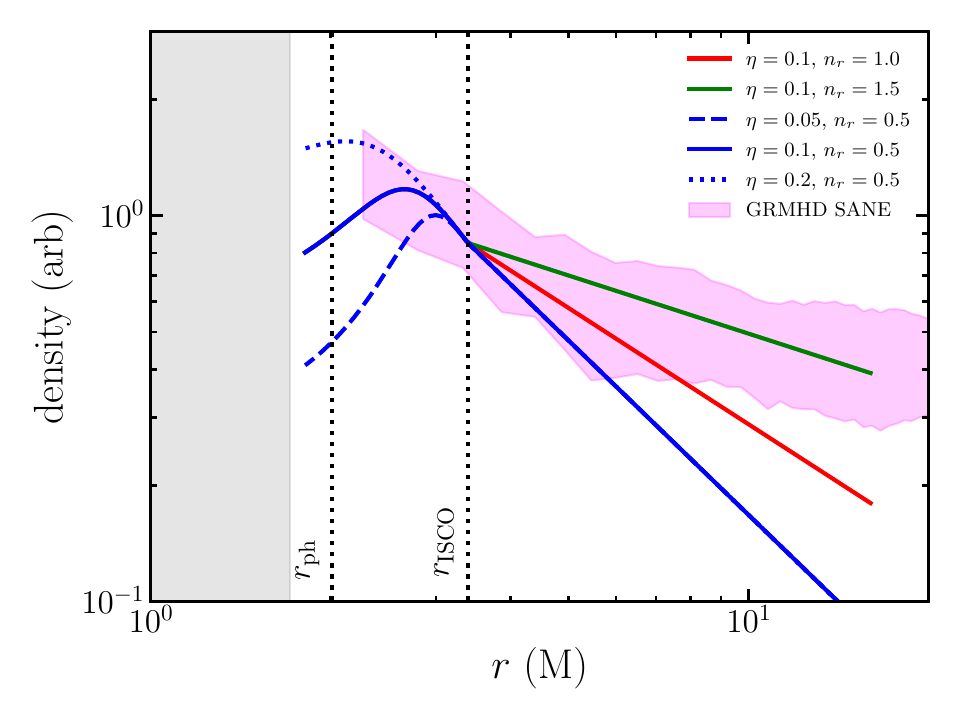}}
 \centerline{
    \includegraphics[width=8.5cm] {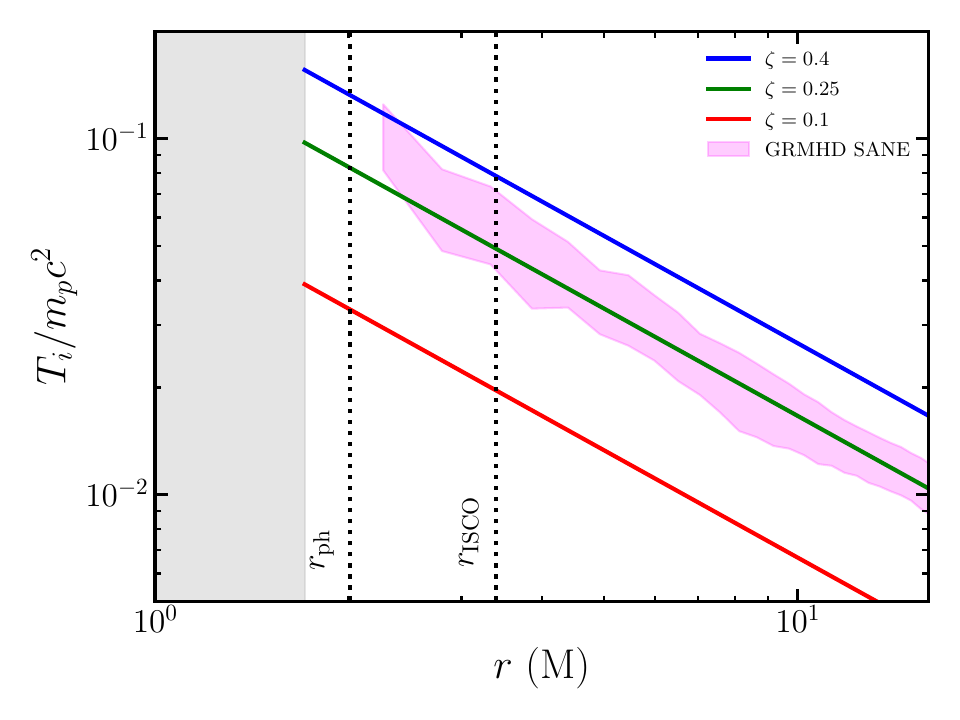}
     \includegraphics[width=8.5cm] {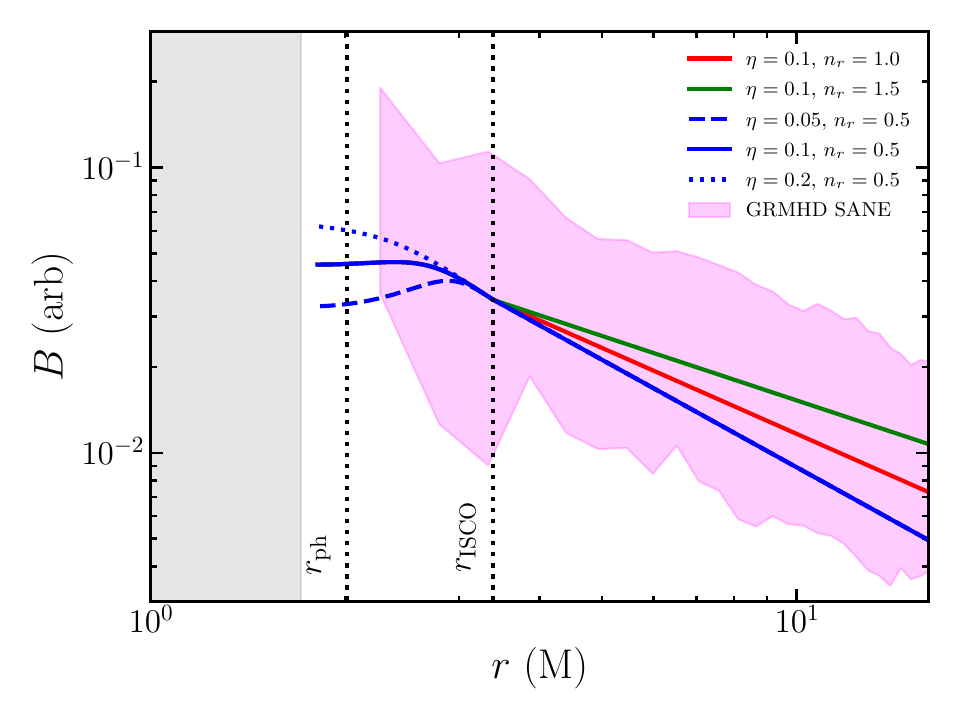}}             
    \caption{\footnotesize Comparison of the radial profiles of various flow properties in the analytic model to an example GRMHD simulation with a black hole spin $a=0.7$. Top left shows the azimuthal velocity, top right the equatorial density, bottom left the ion temperature, and bottom right the magnetic field strength. GRMHD simulation results have been averaged over the azimuthal direction and over multiple snapshots. Analytic models derived from conservation laws show good agreement with the detailed GRMHD models, as expected. } 
    \label{fig:grmhd_comp}
\end{figure*}

\subsection{Can Plasma Cooling Truncate the Emissivity?}

We finally explore one last possibility related to plasma thermodynamics to assess its impact on the emissivity profile in the flow. From energy conservation in eq.~(\ref{eq:Econs}), it is evident that the ion (and electron) temperatures are set by the balance between the rate of viscous and compressional heating and synchrotron cooling in the fluid. In the Appendix, we considered the case where the viscous heating rate may drop rapidly at the ISCO as the shear stresses vanish and the ions and electrons are subject only to compressional heating, neglecting synchrotron cooling. If instead we neglect all sources of heating interior to the ISCO, the electron temperature will then only depend on the synchrotron cooling. We now calculate this cooling timescale and evaluate whether it can have a substantial effect on the electron temperature within the free fall time from the ISCO. 

The frequency integrated synchrotron emissivity is given by 
\begin{equation}
     j_{\rm syn} = \frac{9 \sqrt{3} c \sigma_T n_{\rm{e}} B^2 \theta_e^2}{256 \pi^3 K_2(1/\theta_e)} 
     \int_0^\infty M(x_M) x_M dx_M, 
\end{equation}
with $M(x_M)$ defined in equation~(\ref{eq:mxm}) and $\sigma_T$ is the Thomson cross section. The last integral asymptotes to $\sim 26$ for the temperature and frequency regime relevant for mm-wavelength images of radiatively inefficient accretion flows. 

For a fluid element of relativistic particles at a given temperature radiating at a characteristic magnetic field strength in the inner accretion flow, the cooling timescale can be estimated by
\begin{equation}
    t_{\rm cool} \sim \frac{3 n_e k_B T}{j_{\rm syn}}\;.
\end{equation}
Taking the high-temperature limit of synchrotron emissivity and evaluating it at conditions that are typical for the flows under consideration, we find
\begin{eqnarray}
    t_{\rm cool} &=&
    \frac{256 \pi^3 c^3 m_e^2}{3 \sqrt{3} B^2 \sigma_{\rm T} k_{\rm B} T}
    \left(\int_0^\infty M(x_M) x_M dx_M\right)^{-1}\nonumber\\
    &\simeq & 3 \times 10^6 \left(\frac{B}{10\; {\rm G}}\right)^{-2} \left(\frac{T}{5\times 10^{10} \; {\rm K}}\right)^{-1} {\rm s}.
\end{eqnarray}
We compare this to the free-fall time from the ISCO, which is given by 
\begin{equation}
    t_{\rm ff} = \frac{r^{3/2}_{\rm ISCO}}{(2GM)^{1/2}} = 
    3.3 \times 10^5  \left(\frac{r_{\rm ISCO} \; c^2}{6GM}\right)^{3/2} \left(\frac{M}{6.5 \times 10^9 M_\odot}\right) {\rm s}.
\end{equation}
The ratio of the free-fall time to the cooling becomes 
\begin{eqnarray}
\frac{t_{\rm ff}}{t_{\rm cool}} = 0.1
\left(\frac{M}{6.5 \times 10^9 M_\odot}\right)
\left(\frac{r_{\rm ISCO} \; c^2}{6GM}\right)^{3/2} \\ \nonumber
\times \left(\frac{B}{10\;G}\right)^2
\left(\frac{T}{5\times 10^{10}\;K}\right).
\end{eqnarray}
This implies that, even if electron heating ceases completely at or near the ISCO radius, the particles will fall in \sout{much} faster than they can radiate away their energy. As a result, even in this extreme scenario, the temperature is not expected to change significantly interior to the ISCO. Therefore, a precipitous drop in the emissivity is inconsistent with any fluid model that obeys conservation laws and cannot cause a suppression in the image brightness that is disjoint from the black hole shadow.

\section{Comparison to GRMHD Simulations}

The analytic model presented in the previous sections allow us to calculate the macroscopic properties of the gas in the accretion flow based only on conservation laws and some basic assumptions. This model also shows that the microscopic physics in the plasma does not have a significant influence on these large scale properties and the corresponding brightness profiles from the inner accretion flows for configurations that produce compact ring-like images. 

Addressing more detailed questions about accretion flows, such as jet/wind launching, variability properties, and the fine structure of images, on the other hand, requires the use of numerical simulations, which allow the incorporation of physics of the plasma at different spatial scales. GRMHD simulations provide an avenue for exploring some of this microphysics more broadly, under different sets of assumptions, but at their core are still based on basic conservation laws for fluids. This leads to the expectation that the analytic model should capture the average macroscopic properties of GRMHD simulations. In this section, we use an example GRMHD simulation to demonstrate this point. 

To this end, we utilize the average physical quantities from a GRMHD simulation with a SANE (standard and normal evolution) initial magnetic field configuration and a black hole spin parameter $a=0.7$ that was first introduced in \citet{Narayan2012} and \citet{Sadowski2013}. The accretion flow is evolved using the {\tt HARM3D} code \citep{Gammie2003} from an initial torus located between 10M and 1000M. The simulation was run for a time span of $t=200,000~GM/c^3$ to ensure that steady state is reached in the inner disk. The gas has an adiabatic index $\hat{\gamma}=5/3$. Several subsequent studies explored the images~\citep{Chan2015}, variability~\citep{Chan2015b}, interferometric observables~\citep{Medeiros2017,Medeiros2018}, and the flaring properties~\citep{Ball2016} of these GRMHD simulations. 

In Figure~\ref{fig:grmhd_comp}, we show the radial profiles of the azimuthal velocity, density, ion temperature, and magnetic field strength from the GRMHD simulation and compare them with the analytic models with different parameters we introduced earlier. The pink band for the GRMHD outputs represents the range we obtained by averaging over the azimuthal direction and sampling multiple snapshots from the long simulation that are far apart in time to ensure that the profiles are representative. In order to capture the physical quantities only near the equator, we also average over polar angles within $\pm \pi/8$ from the equator. In all cases, analytic models correctly capture the broad characteristics of the GRMHD profiles, demonstrating that these profiles are determined primarily by the general physical considerations and conservation laws that govern the flows rather than by the details of the plasma processes. 

\begin{figure*} 
 \centerline{
    \includegraphics[width=8.5cm] {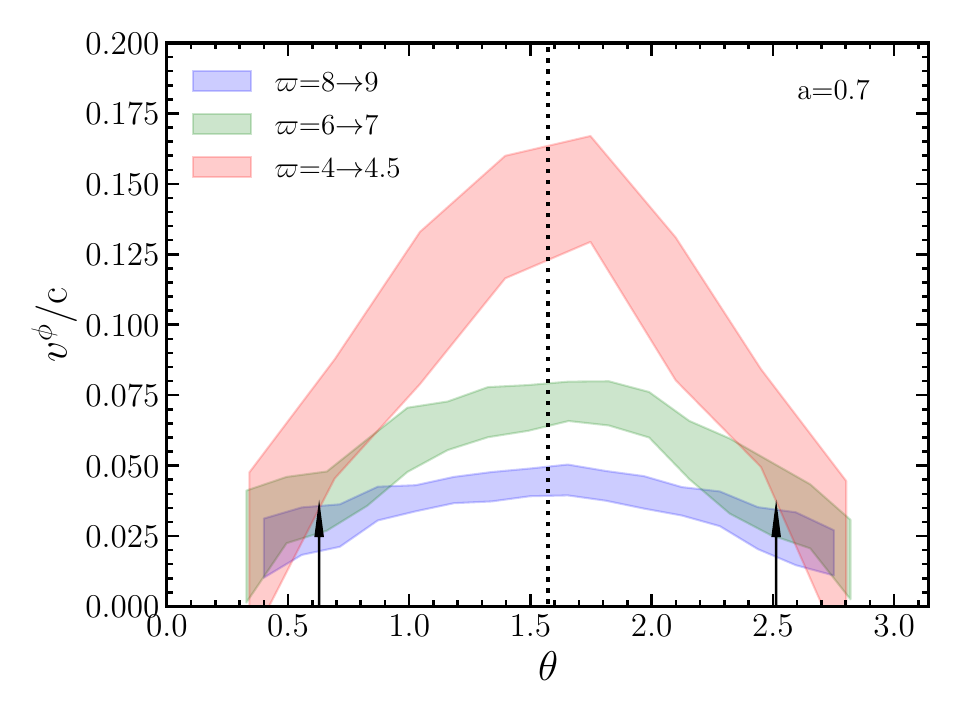}
    \includegraphics[width=8.5cm] {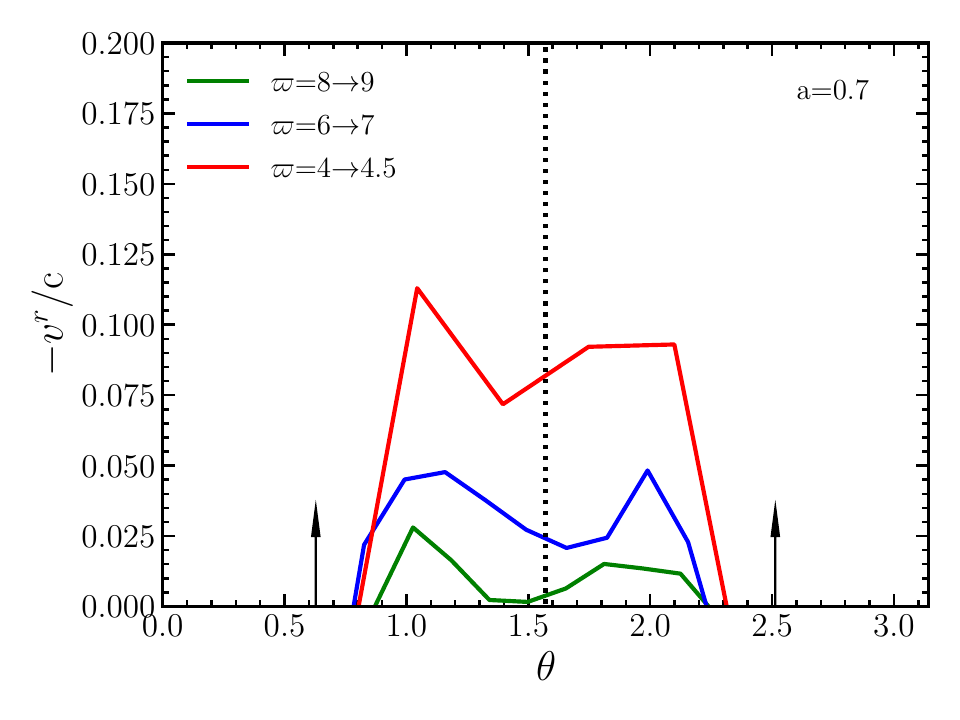}}
 \centerline{
     \includegraphics[width=8.5cm] {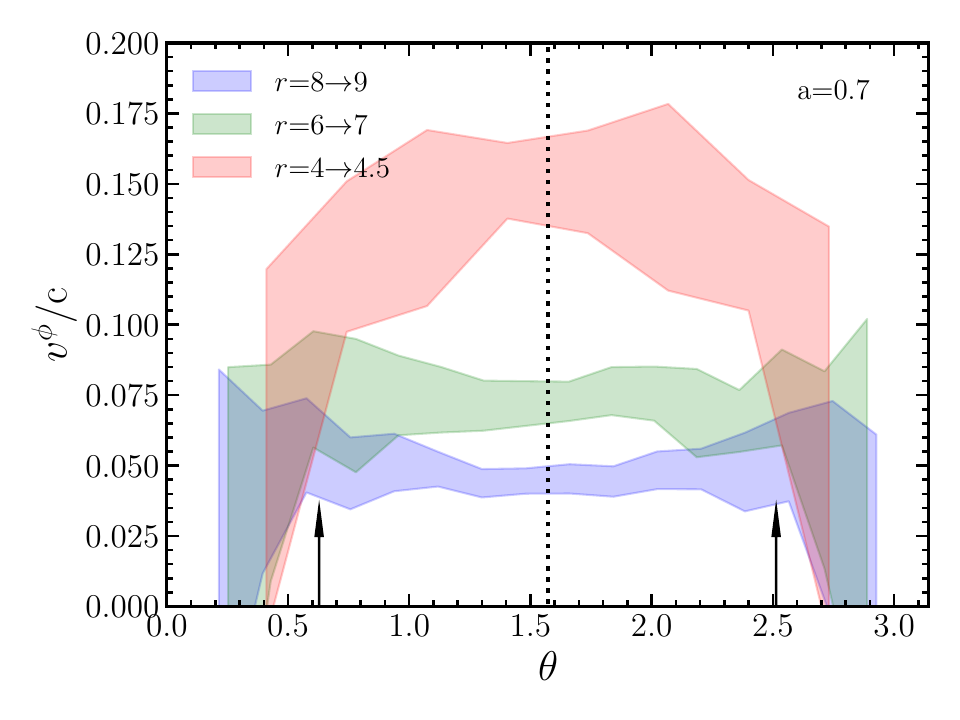}
          \includegraphics[width=8.5cm] {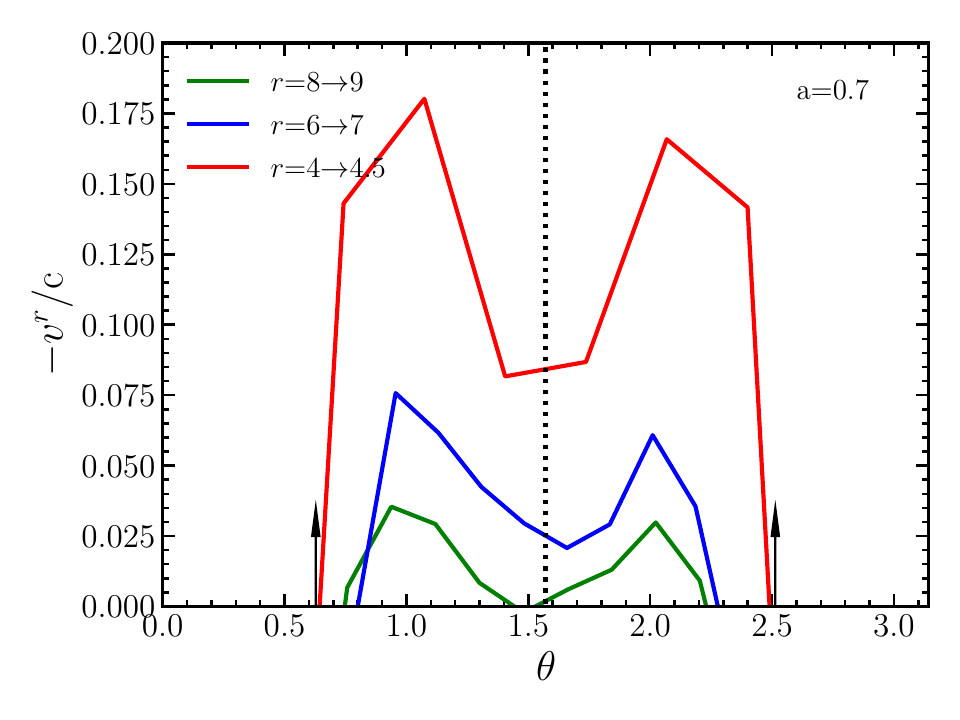}}            
    \caption{\footnotesize Radial and azimuthal components of the velocity in the flow as a function of the polar angle $\theta$. The top panels show the velocity components at constant cylindrical radii $\varpi$ and the bottom panels at constant spherical radii $r$ for a GRMHD simulation with a black hole spin $a=0.7$. The shaded bands in the left panels show the rms range of the azimuthal velocities, for different azimuths and  snapshots. The curves in the right show the mean values of the radial velocities, averaged over azimuth and time. The vertical arrows in all panels denote the polar angle that corresponds to one disk scale height $\theta=h/r$. As expected from the self-similar analytic solutions, the radial velocity $v^r$ is constant along cylindrical surfaces whereas the angular velocity $u^\phi$ is constant on spherical surfaces. 
    }
    \label{fig:velocity_theta}
\end{figure*}

\begin{figure*} 
 \centerline{
    \includegraphics[width=8.5cm] {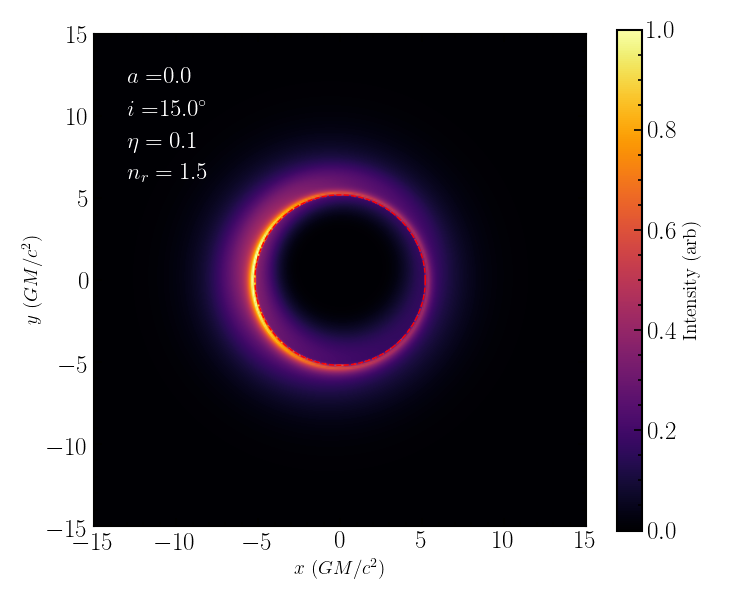}
     \includegraphics[width=8.5cm] {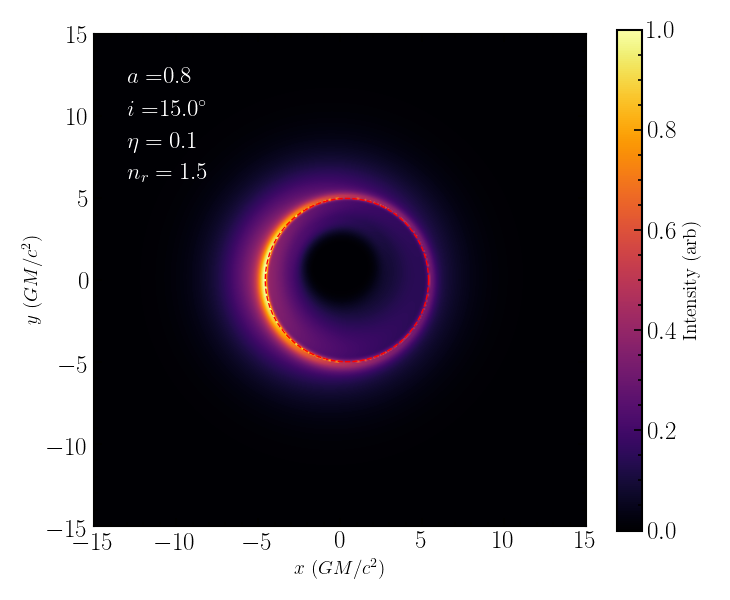}}
 \centerline{
    \includegraphics[width=8.5cm] {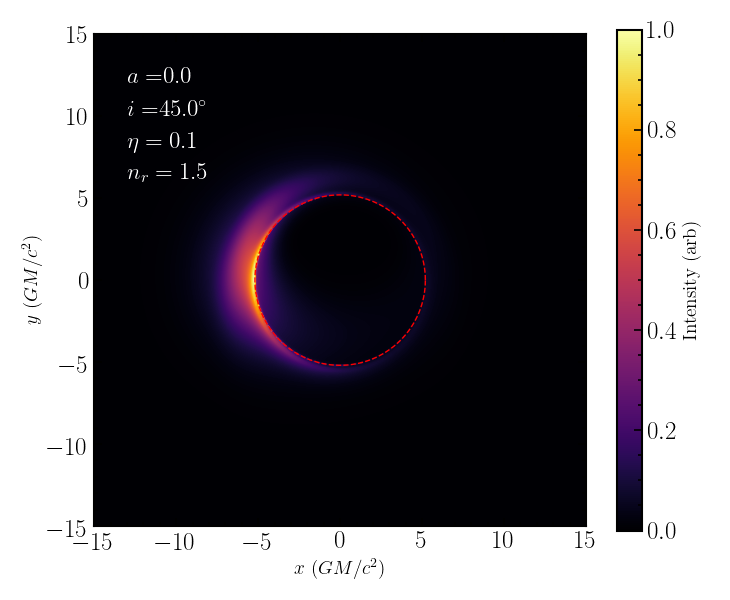}
     \includegraphics[width=8.5cm] {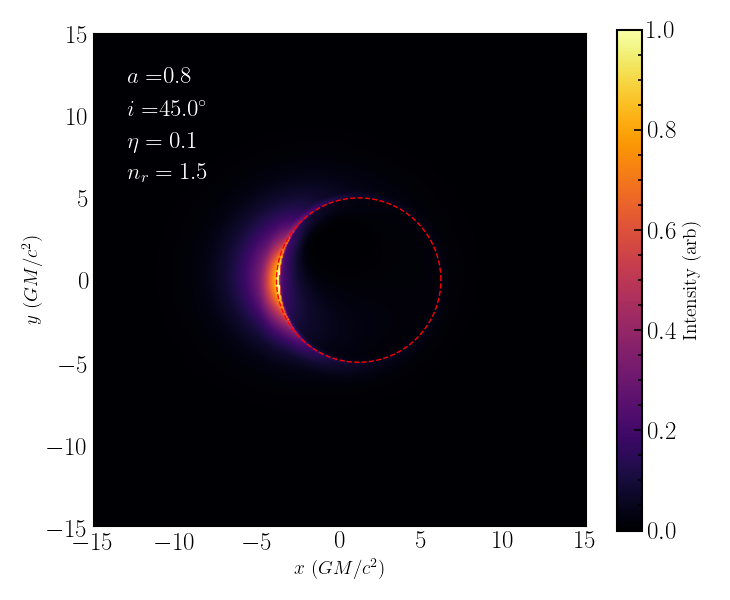}}   
\centerline{
    \includegraphics[width=8.5cm] {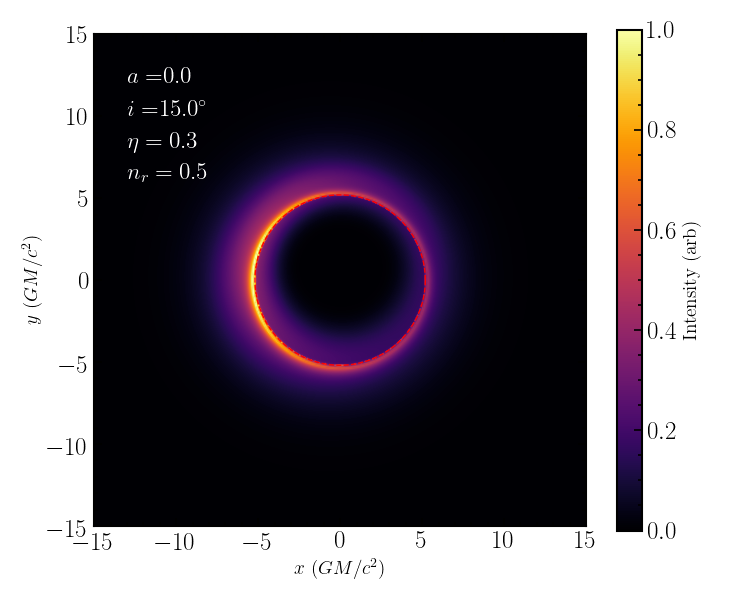}
     \includegraphics[width=8.5cm] {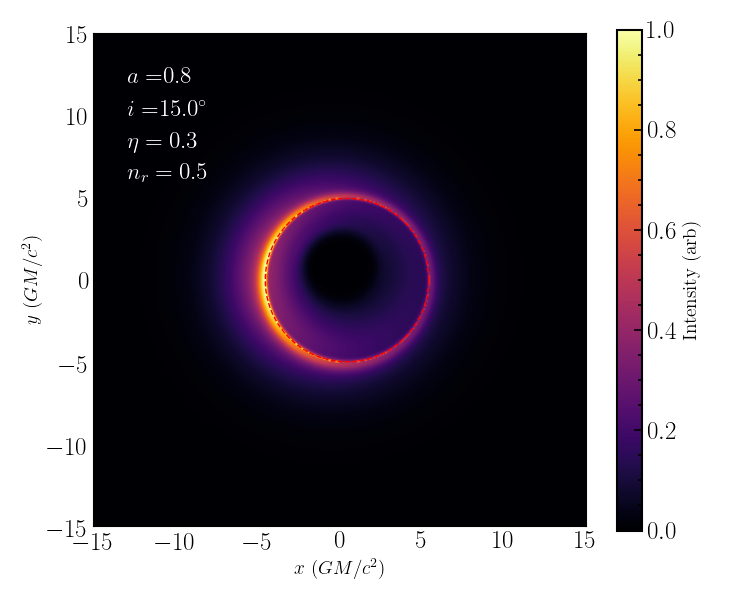}}    
    \caption{\footnotesize Images of black holes at 1.3mm calculated using the analytic plasma model described in \S3, for different black-hole spins, observer inclinations, and parameters of the radial velocity profile. In all panels, we have set $R=5$ and $B_0=20$~G. The outline of the black-hole shadow on each image is shown as a dashed red line. The bright ring of emission closely traces the boundary of the shadow for a wide range of assumptions regarding the spacetime, the plasma properties, or the observer.} 
    \label{fig:images}
\end{figure*}

The azimuthal velocity profile follows the radial dependence of the Keplerian velocity at all radii but is sub-Keplerian everywhere, as is expected for radiatively inefficient flows that have significant pressure support. In the remaining three panels, the density, temperature, and the magnetic field strength monotonically rise inward in the simulations, supporting the conclusions from continuity arguments that accretion flows do not have characteristics that can cause a sudden drop in emissivity outside the event horizon.

\section{Images}

Having developed a covariant analytical model for the height-averaged flow quantities, we turn to calculating black hole images at 230\;GHz. Calculating these images requires specifying the density, temperature, magnetic field, and the flow velocities at all locations in the spacetime, instead of only at the equatorial plane, as we have done so far.  We calculate these quantities off the equatorial plane based on their equatorial values and the following simple physical arguments. 

Because we assumed axisymmetry, a given point in the spacetime is specified by its radius and polar angle, i.e., by $(r,\theta)$. A spherical surface that goes through this point has a constant spherical radius $r$. A cylindrical surface that goes through the same point has a constant cylindrical radius $\varpi=r \sin\theta$.

\noindent {\it Plasma Properties off the Equatorial Plane.---\/} We specify the electron density using the definition of the finite disk scale height $h/r$. In particular, we multiply the equatorial density profile $n_{\rm e, eq}$ with an exponential in the polar angle $\theta$, i.e.,
\begin{equation}
    n_{\rm e}(r,\theta)=n_{\rm e, eq}(\varpi)\exp\left\{-\frac{1}{2}\left[\frac{\theta-\pi/2}{(h/r)\pi/2}\right]^m\right\}, 
\label{eq:den_offeq}
\end{equation}
where the index $m$ determines the slope of the vertical density profile. In the Newtonian limit and for an ion temperature that is constant with height, $m=2$; we will use this value hereafter, unless we specify otherwise. Following eq.~(\ref{eq:ne_gen}), we set the equatorial electron density profile to
\begin{equation}
n_{\rm e ,eq}(\varpi) = \frac{\dot{M}}{4 \pi \varpi^2 (h/r) u^r m_{\rm p}}\; . 
\label{eq:ne_3D}
\end{equation}

In order to be consistent with the above expression for the density profile, we set the electron temperature $T_{\rm e}$ to its equatorial value at the corresponding spherical radius. Using eq.~(\ref{eq:Te}), we write
\begin{equation}
    T_{\rm e}(r,\theta)=\frac{m_p c^2}{k_{\rm B}} \frac{(\hat{\gamma}-1)}{(R+1)} \zeta \left(\frac{GM}{r c^2}\right)\;.
\label{eq:Te_3D}
 \end{equation}

Finally, we specify the magnetic field everywhere such that the plasma-$\beta$ parameter is constant throughout the flow, i.e., such that
\begin{equation}
B(r,\theta) \propto \left[ n_{\rm e}(r,\theta) T_i (r,\theta)\right]^{1/2}\;.
\label{eq:B_beta}
\end{equation}
In this expression, the overall scale of the magnetic field in the accretion flow depends on the accretion rate. However, because we do not consider explicitly the effects of synchrotron self-absorption in the present calculation, the overall normalization of the accretion rate does not enter the calculation explicitly. For this reason, we simply specify the strength of the magnetic field at a fiducial equatorial location and scale it according to relation~(\ref{eq:B_beta}). In other words, we write
\begin{equation}
    B(r,\theta)=B_0 \left[\frac{n_e(r,\theta) T_i(r,\theta)}{n_e(6M,\pi/2) T_i(6M,\pi/2)}\right]^{1/2}\;,
\end{equation}
such that $B_0$ is the strength of the magnetic field at $r=6M$ on the equatorial plane.

\noindent {\it Velocities off the Equatorial Plane.---\/} 
To specify the radial and azimuthal components of the velocity off of the equatorial plane, we turn to the results of semi-analytical models (e.g., Narayan \& Yi 1996) and of GRMHD simulations (e.g., Sadowski et al. 2013) of geometrically thick flows. These self-similar analytic solutions as well as numerical simulations suggest that {\it (i)} the azimuthal component $u^\phi$ of the velocities are approximately constant on spherical surfaces and 
{\it (ii)} the radial component $u^r$ of the velocities are approximately constant on cylindrical surfaces. 

We show an example of this behavior from a GRMHD simulation in Figure~\ref{fig:velocity_theta}, where we plot as a function of the polar angle $\theta$ the radial and azimuthal components of the velocity in the same SANE simulation that was introduced in \S4. Indeed, the top right panel shows that the radial velocity varies weakly with $\theta$ at constant cylindrical radius $\varpi$, while the lower left panel shows that the azimuthal velocity depends weakly on $\theta$ at constant spherical radii. This behavior is typical in other simulations as well. 

Based on these considerations, we write
\begin{equation}
   u^\phi(r,\theta)=u^\phi_{\rm eq}(r) 
\end{equation}
and 
\begin{equation}
    u^r(r,\theta)=u^r_{\rm eq}(r\sin\theta)
\end{equation}
and calculate $u^t$ from the requirement $\vec{u}\cdot\vec{u}=-1$. We note that this model breaks down when $r\sin\theta\le r_{\rm hor}$, where $r_{\rm hor}$ is the equatorial horizon radius of the black hole. We, therefore, excise this cylinder of radius $r_{\rm hor}$ from the simulation. Moreover, this model introduces some pathologies to the velocity profile for near maximum negative black hole spins ($a\lesssim -0.9$) because of the large mismatch between the vertical dependence of the angular velocity of the plasma and of the rate of frame dragging, which occurs in the opposite sense.

Figure~\ref{fig:images} shows six examples of images calculated at 230~GHz by integrating the radiative transfer equation along geodesics from an image plane at infinity at an inclination $i$ from the spin axis of the black hole. In each panel, we vary various spacetime and plasma characteristics. As found in all earlier numerical investigations of black-hole images that are dominated by the accretion flows, the structure of the image is dictated primarily by gravitational lensing and, in particular, by the presence of the deep brightness depression that we identify with the black-hole shadow. This shadow is surrounded by a bright, narrow ring of emission. The boundary of the shadow is always contained within the bright ring. In other words, the bright ring in the image closely traces the boundary of the shadow. Therefore, the diameter of this ring can be used as a proxy for the diameter of the shadow, when an appropriate calibration factor between the two is applied.

\begin{figure*} 
 \centerline{\includegraphics[width=0.9\textwidth] {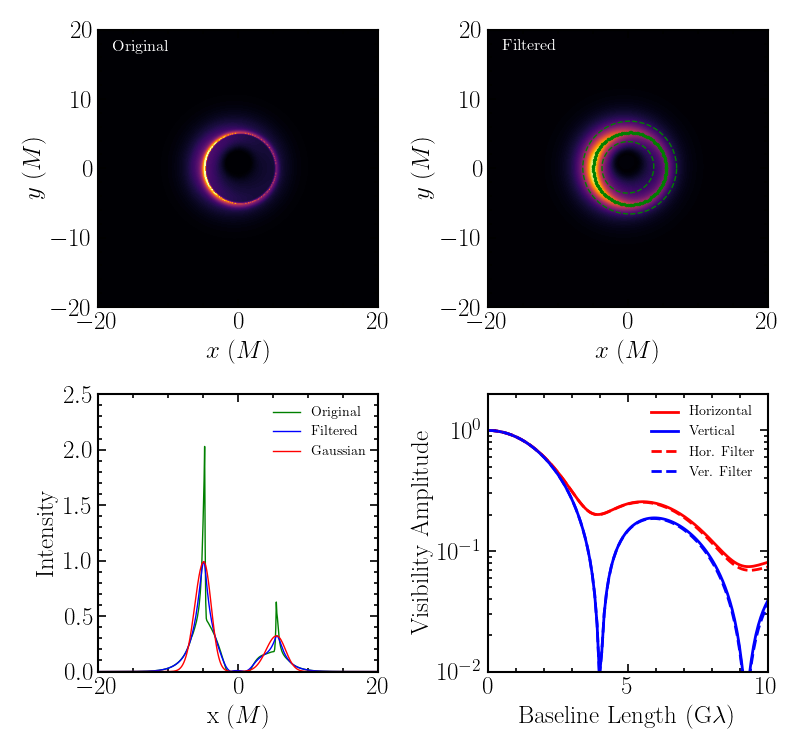}}            
    \caption{\footnotesize A demonstration of the image characterization algorithm employed in \S5. {\it (Upper left)\/} A sample 1.3~mm black-hole image for a spin of $a=0.6$, an observer inclination of $30^\circ$, and the analytic plasma model described in \S3. {\it (Upper right)\/} The same image after a 15~$G\lambda$, $n=2$ Butterworth filter has been applied to mimic the finite resolution of the EHT. The solid green curve identifies the location of maximum brightness along each radial cross section of the image. The dashed green curves identify the FWHM of the brightness distribution of the equivalent Gaussian representation of the image along each radial cross section. {\it (Lower left)\/} A horizontal cross section of the image. The green curve shows the brightness of the original image; the blue curve shows the brightness of the filtered image; the red curve shows the equivalent Gaussian representation of the cross section. {\it (Lower right)\/} Two cross sections of the visibility amplitudes of the original and filtered images. The Butterworth filter only marginally affects the visibility amplitudes and not at the locations of the deep minima, on which the observational measurement of the image size is based.}
    \label{fig:filter}
\end{figure*}

\subsection{Characterizing Black-Hole Images}

We now turn to identifying the location of the peak brightness of a black-hole image and quantify its relation to the size of the black-hole shadow. In principle, as shown in all previous examples, the peak brightness always occurs at the critical impact parameter because, be definition, the optical path along the critical null geodesics is infinite. However, the finite resolution of the EHT does not allow us to resolve this photon ring on the black-hole image, which appears blended with the nearby emission (see, e.g., Fig.~\ref{fig:images}). 
For this reason, we have developed an image-domain characterization algorithm that allows us to quantify the size of the bright ring in a black-hole image in a manner that accommodates the finite resolution of the EHT but, at the same time, does not alter the interferometric observables on which the measurement is based.

The image-domain characterization algorithm comprises the following steps (see Fig.~\ref{fig:filter} for a visual representation of the algorithm):

\noindent {\it (i)\/} We first filter the original image at the nominal resolution of the EHT. Following Psaltis et al. (2021), we employ a Butterworth filter with $n=2$ and a characteristic scale of 15~$G\lambda$. Unlike a Gaussian filter, the Butterworth filter minimizes any alteration of the image at scales that are accessible to the EHT observations, while suppressing any small-scale structures that are not.

\noindent {\it (ii)\/} We calculate analytically the center of the black-hole shadow. This is displaced from the coordinate center of the image because of the effects of differential frame dragging. For the Kerr metric we employ here, this displacement depends only on black-hole spin and inclination.  

\noindent {\it (iii)\/} Starting from the center of the black-hole shadow, we use a rectangular bivariate spline interpolation to obtain radial cross sections of the filtered image brightness at 128 equidistant azimuthal orientations. We define the fractional coverage of a ring-like shape ${\cal F}$ as the fraction of these radial cross sections for which the image brightness is at least $10\%$ of the maximum of the entire image. We chose this value to reflect the dynamical range of $\sim 10$ of the 2017 EHT images. \textbf{}

\noindent {\it (iv)\/} We measure, in each radial cross section, the distance of the location of peak brightness from the center of the black-hole shadow. We identify the diameter of the bright emission ring as twice the median value of this distance.

\noindent {\it (v)\/} For each radial cross section, we generate an equivalent asymmetric Gaussian representation of the brightness by setting the location and peak brightness of the Gaussian equal to those of the filtered cross section and the widths of the asymmetric Gaussian towards larger and small radial distances such that the corresponding integrated brightness of the cross section of the filtered image is equal to that of the Gaussian. We measure the FWHM of the asymmetric equivalent Gaussian representation for each cross section and identify their median with the FWHM of the bright ring. 

\begin{figure*} 
 \centerline{
    \includegraphics[width=0.5\textwidth] {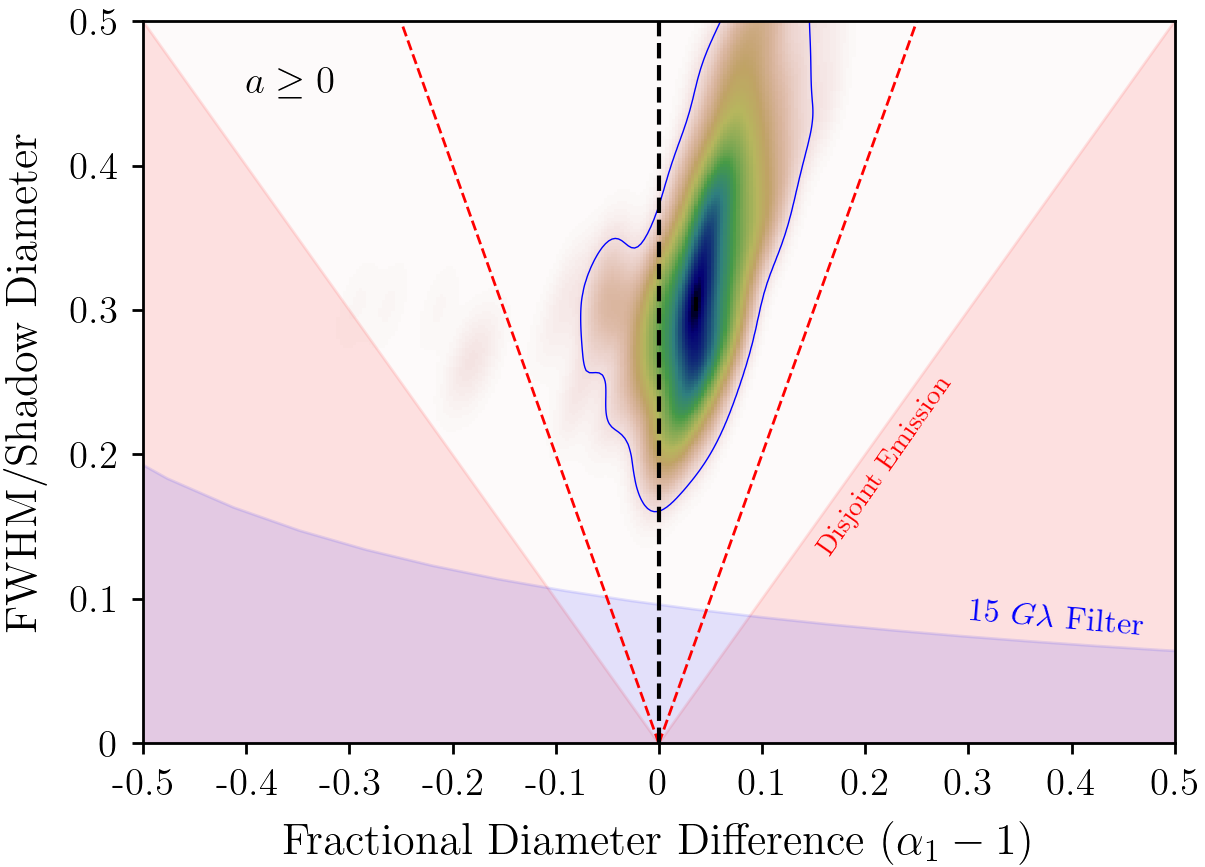}
       \includegraphics[width=0.5\textwidth] {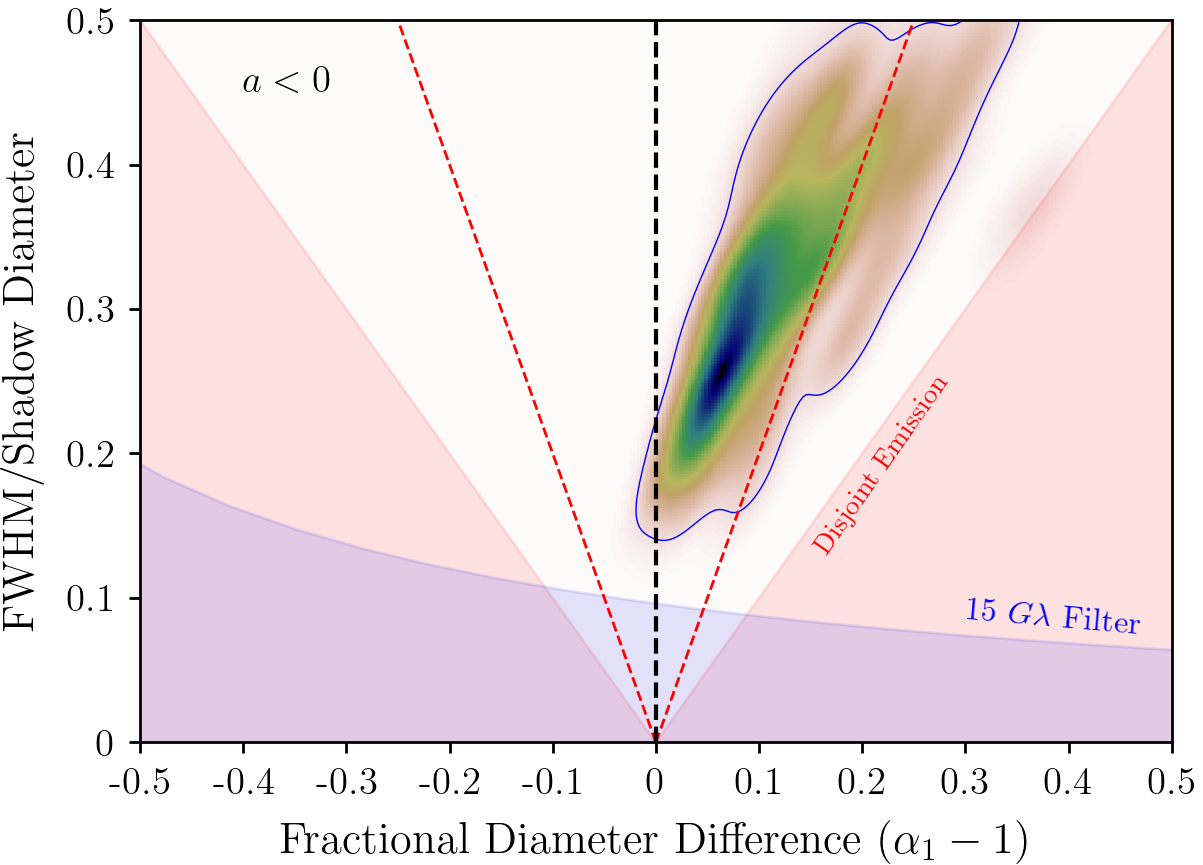}} 
    \caption{\footnotesize The fractional difference $\alpha_1-1$ between the average diameter of peak brightness in a 1.3~mm black-hole image and the diameter of the black-hole shadow (i.e., twice the critical impact parameter) shown against the fractional FWHM of the bright ring in the image, for a broad range of black-hole spins, observer inclinations, and plasma model parameters (see text for details). The left and right panels correspond to positive and negative black hole spins, respectively. The purple-filled area shows the minimum fractional width of the ring imposed by the $15~G\lambda$ Butterworth filter that has been applied to the image to mimic the finite resolution of the EHT. The red-filled area represents configurations in which the fractional diameter difference is larger than the fractional FWHM of the bright ring; the dashed red line corresponds to the fractional diameter difference being equal to the fractional HWHM of the bright ring. If configurations existed in the red-filled area, their black-hole shadows would have been disjoint from the bright rings. All the images fall near the vertical dashed line at $a_1-1=0$, demonstrating that, in all cases, the diameter of the black-hole shadow is very close to that of peak brightness. The fact that all images fall within the white area demonstrates that the bright rings always encompass the critical impact parameters.}
    \label{fig:bias}
\end{figure*}

Employing this algorithm, we measure the diameter $d_{\rm im}$ and FWHM of the bright emission ring in each simulation image and compare them to the average diameter $d_{\rm sh}$ of the Kerr black-hole shadow. For the latter, we use the analytic approximation derived in \citet{Chan2013}
\begin{equation}
    d_{\rm sh}=2 R_0 +2 R_1 \cos(2.14 i -22^\circ.9)\;,
    \label{eq:dGR}  
\end{equation}
where the inclination $i$ is expressed in degrees and
\begin{eqnarray}
R_0 &=& (5.2-0.209 a +0.445 a^2 -0.5673 a^3)M\nonumber\\
R_1 &=& \left[0.24-\frac{3.3}{(a-0.9017)^2+0.059}\right]\times 10^{-3} M\;.
\end{eqnarray}
In particular, we define the ratio $\alpha_1$ between the two diameters as
\begin{equation}
    \alpha_1 \equiv \frac{d_{\rm im}}{d_{\rm sh}}
\end{equation}
and the fractional width of the ring as FWHM/$d_{\rm GR}$. We will refer to the difference $\alpha_1-1$ as the fractional diameter difference. If the diameter of the image is equal to the diameter of the shadow, then $\alpha_1-1=0$. Note that calibrating the image diameter inferred from observations to the shadow diameter has a second component ($\alpha_2$) that quantifies potential biases introduced by the imaging and model-fitting algorithms. To distinguish the purely theoretical displacement explored here from the observational component, we refer to the former as $\alpha_1$.

Using this method, we can quantify the expected fractional diameter difference between the diameter of the bright image ring and that of the shadow for a very broad range of model, black-hole, and observer parameters. Figure~\ref{fig:bias} shows the fractional diameter difference $\alpha_1-1$ and the fractional FWHM measured from images that use the semi-analytic plasma model described in \S3. We employed a grid of model images in which the black-hole spin was varied as $a= -0.65, -0.32, 0.0, 0.29, 0.56, 0.78, 0.94$, corresponding to equidistant spacing in ISCO radii. The observer inclination was varied as $i=15, 30, 45, 60, 75$ degrees. We set the scale of the magnetic field to $B_0=5$~G, $20$~G, and $50$~G, based on theoretical expectations~\citep{Satapathy2021} and limits arising from the observed polarization signatures in the M87 image~\citep{PaperVIII}. We set the ion-to-electron temperature ratio to $R=5$ and 10. We do not consider the case $R=1$, which is inconsistent with the assumption of a radiatively inefficient flow (see \S6) or larger values  of $R$ for which the emission from the accretion flow, which we model here, is eliminated. Finally, we take the following two sets of parameters for the radial velocity profile: $\eta=0.05,  0.1, 0.2$ for $n_r=1.5$ and $n_r=0.5, 1.0, 1.5$ for $\eta=0.1$. We then characterized all 3780 images using the algorithm described above and consider in this figure all images for which ${\cal F} \geq 0.5$, such that a radius can be defined. 

The blue shaded region in the figure corresponds to the FWHM that an infinitesimal ring would have been broadened to by our 15~$G\lambda$ filter. In order to apply this filter, we have assumed that the angular size of one gravitational radius for a black hole located at distance $D$ is $\theta_{\rm g}\equiv GM/(c^2 D)=3.6 \mu$as, i.e., similar to that of the M87 black hole.

In each of the simulated images, the diameter of the ring that the EHT would observe is comparable to and, in general, only marginally larger than the diameter of the black-hole shadow. This is expected from the preceding discussion, which presented the physical reasons why the image diameter is not identically equal to that of the shadow, i.e., in general,  $\vert\alpha_1-1\vert \neq 0$. Furthermore, the exact value of the fractional diameter difference does depend on the plasma model, but it is always a small correction. As this figure shows, the fractional diameter difference is $\vert\alpha_1-1\vert<0.1$ for the positive spins and $\vert\alpha_1-1\vert \lesssim 0.3$ for negative spins.  The higher values of this displacement always correlate with the larger values of the ring widths. In both cases, the peak of the distribution is in good agreement with the value of $\sim 9$\% inferred for a series of 100 GRMHD snapshots in \citet{PaperVI}, albeit the latter also incorporates biases introduced by the model fitting process to the EHT data. 

Perhaps more importantly, the bright rings in the images always encompass the boundaries of the black-hole shadows. This is demonstrated by the fact that no images in Figure~\ref{fig:bias} are within the red-shaded area, the boundary of which is determined by the condition that the fractional diameter difference is equal to the fractional FWHM of the image. In other words, all black-hole images calculated here obey
\begin{equation}
    \vert\alpha_1-1\vert \ll  \frac{\rm FWHM}{d_{\rm GR}}\;.
\end{equation}
Given the definition of the diameter bias, this implies that
\begin{equation}
    d_{\rm GR}-{\rm FWHM}
    \ll d_{\rm im} \ll
    d_{\rm GR}+{\rm FWHM}\;.
\end{equation}
In words, the above inequality describes the fact that, in all images calculated here, the boundary of the black-hole shadow is never disjoint from the bright emission ring that the EHT observes and the expected offset between the two is contained within the measured width of the ring image.

\begin{figure*} 
 \centerline{
    \includegraphics[width=8.5cm]{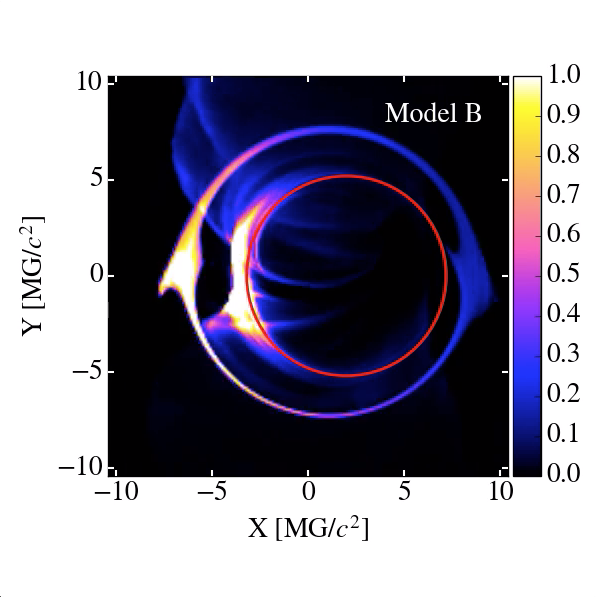}
     \includegraphics[width=8.5cm]{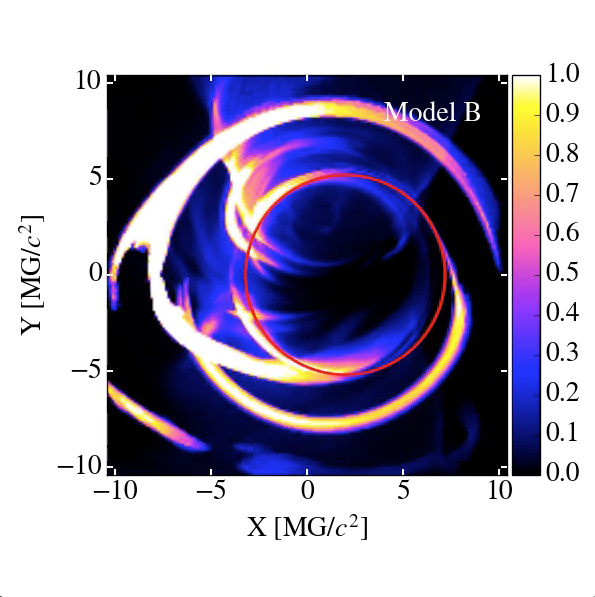}}  
    \caption{\footnotesize Sample snapshots from a SANE GRMHD simulation of accretion around a black hole with spin $a=0.9$, showing two transient events that generate ring-like structures that are disjoint from the black-hole shadow (after \citealt{Chan2015b}). In both cases, the boundary of the shadow is displayed with a red curve. The left panel shows an Einstein ring generated when a very localized structure in the accretion flow crosses a caustic behind the black hole. The right panel shows a structure generated by an azimuthally stretched flux tube that has been lensed above and below the equatorial plane. Both structures, albeit plausible, are very short lived and inconsistent with the inferred stability of the black-hole image at the center of the M87 galaxy.} 
    \label{fig:image_physics}
\end{figure*}

\section{Discussion}

We developed an analytic model based on conservation equations and simple thermodynamic arguments in order to disentangle the effects of the black-hole spacetime imprinted on black-hole images, through the gravitationally lensed optical paths, redshift, and plasma velocities, from those of the emissivity and thermodynamic properties of the plasma. We showed that there is a tight relationship between the location of the bright ring that is observable near the horizon of a black hole and that of its shadow. We further quantified the ratio between the diameter of peak brightness of the image rings and the shadow diameter in order to facilitate the use of millimeter image characteristics observed with the EHT as tests of the black hole metric. 

Our analytic model naturally involves a number of simplifications related to electron physics, such as the use of a constant temperature ratio $R$ between the electrons and ions. However, improving on these simplifications will not change our main qualitative conclusion that the ring-like images always encompass the outline of the shadow. The reason lies on the fact that the photons that create the millimeter images of the EHT targets originate in a very narrow range of distances from the black hole, typically within a few gravitational radii from the horizon. As a result, even though the temperature ratio between the electrons and ions might evolve significantly throughout the flow, only its limited range of values very close to the horizon will affect the image.

The question remains as to whether the models considered here are general enough to definitively support these conclusions. For example, some recent work~\citep{Gralla2019,Gralla2020b} made use of geometrically thin ($h/r \rightarrow 0$) configurations to argue about the characteristics of image formation and the relationship between the diameter of peak brightness and that of the black-hole shadow. Similar calculations were performed earlier for the case of geometrically thin, optically thick accretion disks~\citep{Luminet1979} as well as more recently~\citep{Glampedakis2021}. Albeit instructive as exercises, it is important to acknowledge that such constructions are not applicable to the EHT targets and are not useful in understanding their image properties. Indeed, it is well established that the observed spectral properties and the low inferred radiative efficiencies of these sources require that the accretion flow is optically thin at all but the longest wavelengths. This can be achieved only if the accretion flow consists of a two-temperature plasma in which the ions heat up to the local virial temperature but the electrons cannot couple efficiently to the ions on relevant timescales in order to cool them and radiate away the accretion luminosity (see~\citealt{Narayan1994,Narayan1995a,Narayan1995b}). The large temperature of the ions is what provides the pressure support that makes the accretion flow geometrically thick, rarefied, and optically thin (see eq.~[\ref{eq:hr}]). In such a configuration, the formation of the black-hole image cannot be understood in terms of the lensed images of the accretion disk surface but rather in terms of the optical paths traversed by the photon trajectories that reach the distance observer, as we have done here. 

Perhaps more revealing is the fact that the alternative of a geometrically thin accretion disk (as modeled, e.g., in \citealt{Luminet1979} and \citealt{Gralla2020b}) would not only have been inconsistent with the overall observations of M87, but horizon-scale imaging would not have been possible at all at the millimeter wavelengths used for the EHT observations. A geometrically thin accretion flow (such a Novikov-Thorne disk) would be optically thick and emit nearly blackbody radiation at horizon scales at a much smaller temperature. Obtaining an image similar to that observed by the EHT would have required observations at UV wavelengths\footnote{Such models might be useful if, in the future, new interferometers are developed operating at the UV to X-ray wavelengths necessary to observe horizon-scale images from high-luminosity sources such as quasars~(see, e.g., \citealt{Ozel2001}).}.  

A more plausible situation in which a bright emission ring may be observed that is disjoint from the black-hole shadow is related to transient events, as discussed in~\cite{Chan2015b}. Figure~\ref{fig:image_physics} shows two example snapshots of such events from a SANE GRMHD simulation of a black-hole with spin $a=0.9$ (Model B in \citealt{Chan2015}). In the left panel, an Einstein ring that is clearly displaced from the boundary of the black-hole shadow is formed by lensing when a localized hot flux tube crosses a caustic behind the black hole. In the right panel, a hot flux tube becomes azimuthally sheared by the differential rotation of the flow and appears gravitationally lensed above and below that black-hole shadow. At the resolution of the EHT, this snapshot would also appear as a bright emission ring, but one that is disjoint from the black-hole shadow. 

Even though such transient events appear in simulations and are expected to happen in nature, albeit very rarely because of the alignment required, their distinctive characteristic is temporariness. Their appearance will change dramatically and they will even disappear at timescales comparable to the dynamical timescale in the inner accretion flow. For the black hole at the center of the M87 galaxy, this timescale is $\sim 4-34$~days, depending on the unknown black-hole spin. Nevertheless, there is only evidence for marginal change in the black-hole image across the 7 days of the 2017 EHT observations, which would be inconsistent with such a transient event~\citep{PaperI,Satapathy2021}. More importantly, reanalysis of 1.3~mm data obtained between 2009 and 2017 demonstrate that the images are consistent with a persistent asymmetric ring of $\sim 40~\mu$as diameter that only shows position-angle wandering over a period of a decade~\citep{Wielgus2020}. Further EHT observations separated by $\sim$year timescales can help rule out the possibility of the bright ring image being associated with transient events even more definitively. 

\begin{acknowledgements}

D.\;P.\ and  F.\;\"O.\  acknowledge support from NSF PIRE award OISE-1743747, NSF AST-1715061, and NASA ATP award 80NSSC20K0521.
Z.\,Y.~is supported by a UK Research \& Innovation (UKRI) Stephen Hawking Fellowship and acknowledges partial support from a Leverhulme Trust Early Career Fellowship.
We thank Monika Mo\'scibrodzka and Ramesh Narayan for detailed comments on the manuscript. We also thank Lia Medeiros, Mariafelicia de Laurentis, and all members of the Gravitational Physics Working Group of the EHT, for helpful discussions and comments. We thank the anonymous referee for a careful reading of the manuscript and valuable comments.  
This research has made use of NASA's Astrophysics Data System.

\end{acknowledgements}

\appendix 

\section{Plasma Heating}

In this Appendix, we calculate numerically the integral in eq.~(\ref{eq:Ti_dis}) using one explicit form of the dissipation function in the accretion flow around a Kerr black hole. We then generalize this calculation and discuss the motivation for the analytic form we use in the paper. 

The dissipation function, calculated at the local comoving frame with the plasma, is equal to 
\begin{equation}
\Phi = -t^{(a)(b)} \sigma_{(a)(b)}   = -2 t_{(r)(\phi)} \sigma_{(r)(\phi)}, 
\end{equation}
where we used the fact that the metric in the comoving frame is locally Minkowski and the only non-negligible component of the stress is $r \phi$. The $t_{(r)(\phi)}$ component of the stress tensor in the comoving frame is related to the mixed component in the coordinate frame $t^{r}_\phi$ by~\citep{Gammie2003}
\begin{equation}
    t^{r}_\phi = \gamma r ({\cal A} {\cal D})^{1/2} t_{(r)(\phi)}, 
    \label{eq:transform}
\end{equation}
where $\gamma\simeq 1$ is the local Lorentz factor, 
\begin{equation}
    {\cal A} = 1+a^2/r^2 + 2a^2/r^3
\end{equation}
and 
\begin{equation}
    {\cal D} = 1-2/r + a^2/r^2\;.
\end{equation}

We obtain the stress-energy tensor from the conservation of angular momentum. Using the azimuthal Killing vector $\xi_{(\phi)} = (0,0,0,1)$, we write this as 
\begin{equation}
    \left(T_\mu^\nu \xi_{(\phi)}^\nu\right)_{;\nu} = 0, 
\end{equation}
which, after appropriate averaging and vertical integration, gives
\begin{equation}
    \frac{d}{dr} \left[\dot{M} L_z - 4 \pi \left(\frac{h}{r}\right) r^2 t^r_\phi\right] =0, 
\end{equation}
or equivalently
\begin{equation}
   \dot{M} L_z - 4 \pi \left(\frac{h}{r}\right) r^2 t^r_\phi = \dot{M} j, 
\end{equation}
where j is an eigenvalue of the problem. Under the assumption that the viscous torques vanish at the ISCO, the eigenvalue is equal to the angular momentum at $r_{\rm ISCO}$. However, both numerical and analytic models show that there are non-zero stresses at the ISCO~\citep{Krolik2005,Shafee2008}. To account for both possibilities, we write in general 
\begin{equation}
    j= \lambda L_z(r_{\rm ISCO}). 
\end{equation}
Combining these equations, we obtain
\begin{equation}
    4 \pi \left(\frac{h}{r}\right) r^2 \frac{\Phi}{\dot{M}} = -\frac{2\sigma (L_z-j)}{r ({\cal A}{\cal D})^{1/2}}
\end{equation}

\begin{figure*} 
 \centerline{
    \includegraphics[width=8.5cm] {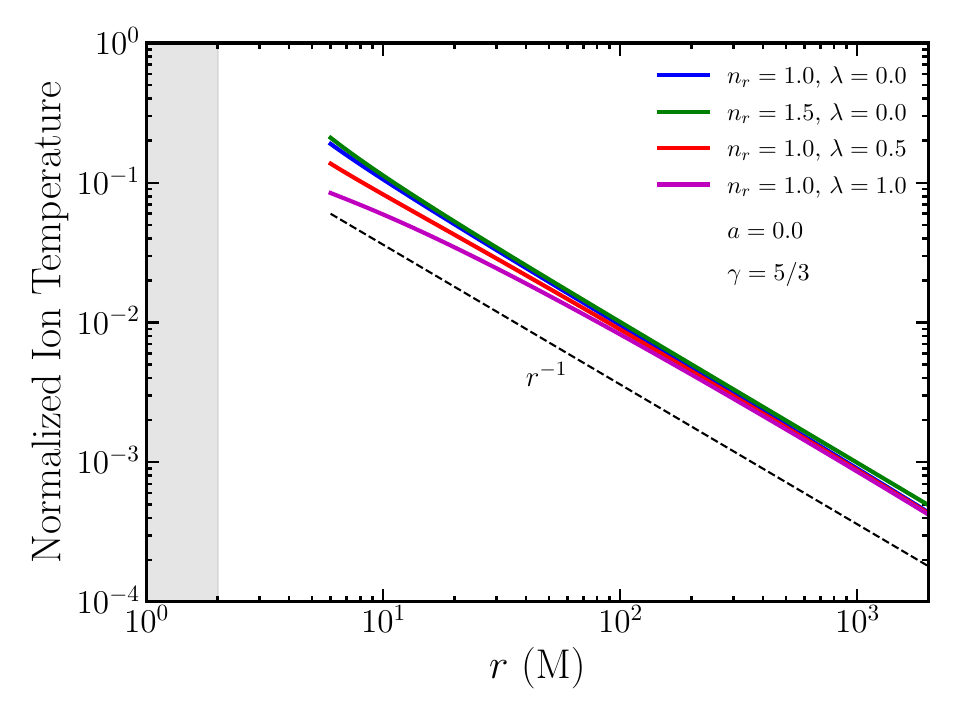}
     \includegraphics[width=8.5cm] {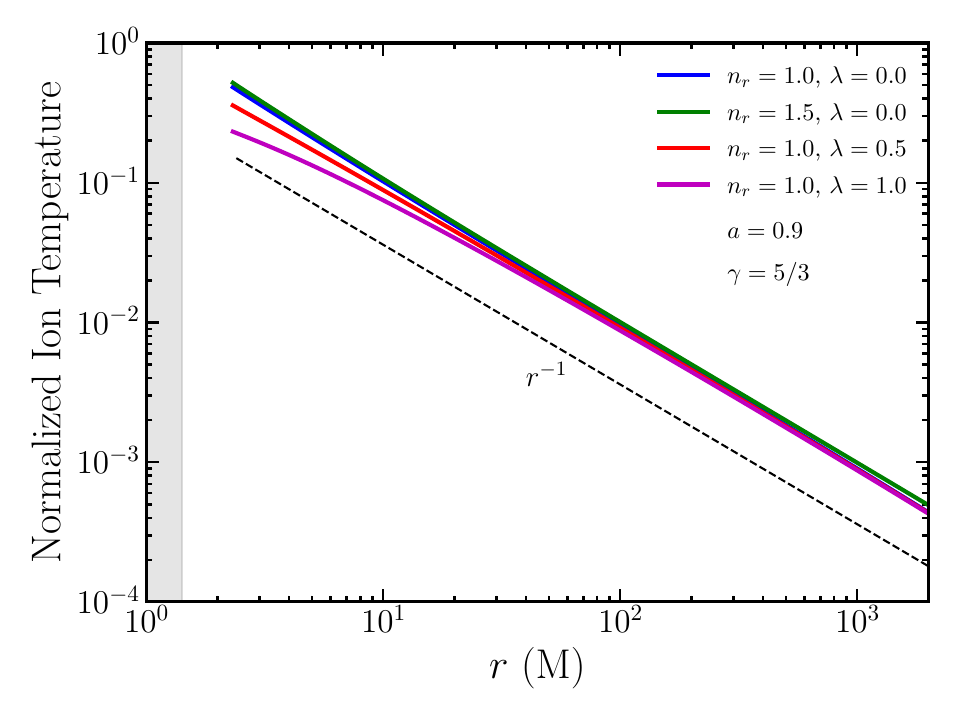}}  
    \caption{\footnotesize The radial profile of the ion temperature for radiatively inefficient accretion flows around Kerr black holes, for different values of the model parameters. Here $n_r$ is the power-law index of the radial velocity profile and $\lambda$ is the angular momentum eigenvalue of the solution. The ion temperature has been normalized by the factor in the square brackets in eq.~(\ref{eq:kT_Newton}) in order to highlight the effects of relativity and the inner boundary conditions. The temperature in a radiatively inefficient flow increases inwards because of the combined effects of ``viscous'' and compressional heating.} 
    \label{fig:Ti_prof}
\end{figure*}

For a geometrically thin flow, where the radial pressure gradients are negligible, we can write (Novikov \& Thorne 1973)
\begin{equation}
    \sigma_{(r)(\phi)} = \frac{1}{2} r {\cal A} \frac{d\Omega}{dr}= -\frac{3{\cal D}}{4{\cal C}} r^{-3/2}\;,
    \label{eq:sigma}
\end{equation}
where 
\begin{equation}
    {\cal C} = 1-3/r + 2a/r^{3/2}. 
\end{equation}
The pressure gradients are, in general, not negligible in radiatively inefficient flows, leading to both sub-Keplerian orbital velocities as well as smaller shear. To account for this, we write 
\begin{equation}
  4 \pi \left(\frac{h}{r}\right) r^2 \frac{\Phi}{\dot{M}} = 
  \frac{3 \epsilon}{2} \left(\frac{{\cal D}}{{\cal A}{\cal C}}\right)^{1/2} (L_z-j) r^{-5/2}\;,
  \label{eq:kT_Kerr}
\end{equation}
where the parameter $\epsilon$ accounts for the non-Keplerian profiles. Using this as well as the radial velocity profile in eq.~(\ref{eq:vr}) and the corresponding density from mass continuity, we can now evaluate the density-weighted integral of eq.~(\ref{eq:calV}). 

In the Newtonian limit, this becomes 
\begin{equation}
    \frac{kT_i}{m_pc^2} = \frac{GM}{rc^2} \left[\left(\frac{R}{R+1}\right) \frac{3 \epsilon (\hat{\gamma}-1)}{6+2 n_r (\hat{\gamma}-1)-4\hat{\gamma}}\right].
    \label{eq:kT_Newton}
\end{equation}
It is important to note that the values of the various parameters in the square bracket cannot all be chosen independently because the radial velocity profile, the deviation from a Keplerian azimuthal velocity profile, and the adibatic index that determines the pressure profile are all coupled to one another through, e.g., the Bernoulli equation. However, regardless of the specific values, the thermal energy in the flow at a given radius is proportional to the gravitational potential energy dissipated down to that radius, as expected. 

Figure~\ref{fig:Ti_prof} shows the ion temperature calculated from eq.~(\ref{eq:Ti_int}) and the complete expression~(\ref{eq:kT_Kerr}) for a Kerr black hole, for two values of the black hole spin parameter. In this figure, the temperature has been normalized by the factor in the square brackets in eq.~(\ref{eq:kT_Newton}) in order to highlight the effects of relativity and the inner boundary conditions. The power-law index of the radial velocity profile, $\eta_r$, determines the radial dependence of the density, via eq.~[\ref{eq:cont}]. Independent of the density profile, the black-hole spin, or the adiabatic index, the $r^{-1}$ temperature profile is sustained down to small radii, with small differences that arise primarily from the particular choice of the angular momentum eigenvalue $\lambda$. The temperature continues to rise inward in all cases, again as expected from energy conservation. 

Inside the ISCO, if there is no additional viscous dissipation, the temperature can either remain constant or increase due to compressional heating. However, extensive numerical and analytic work again shows that the MHD turbulence does not abruptly decay inside the ISCO, and therefore, in realistic set ups, the temperature continues to rise (see Fig.~\ref{fig:grmhd_comp}). 

All the above suggest that the radial velocity structure and corresponding density profile of the accretion flow, as well as the spin of the black hole introduce complexities that are subdominant. This happens because, in a radiatively inefficient flow, the ion temperature at a given radius is determined by the total amount of heat dissipated outwards of that radius, which itself is dictated by the available gravitational potential energy. 

We can now use this understanding to write a general expression that captures the basic properties of ion heating in a general spacetime. We first neglect the subdominant effects of black hole spin and understand all the equations below to be evaluated on the equatorial plane. The transformation~(\ref{eq:transform}) involves a Lorentz boost (for which we assume that $\hat{\gamma}\simeq 1$) and a transformation between the coordinates in which the metric is expressed and those of the local comoving frame. For the latter, we write (see also~\citealt{Bardeen1972})
\begin{equation}
    t^{r}_\phi\simeq \sqrt{\frac{g_{\phi\phi}}{g_{rr}}} t_{(r)(\phi)}\;.
\end{equation}
Similarly, we approximate equation~(\ref{eq:sigma}) by
\begin{equation}
    \sigma_{(r)(\phi)}\simeq \frac{1}{2}\sqrt{g_{\phi\phi}} \frac{d\Omega}{dr}
\end{equation}
and assume that the eigenvalue in the problem is negligible ($\lambda=0$). 

Following the above set of steps, we can then write for the dissipation integral
\begin{equation}
 \int_\infty^r \frac{\Phi}{\dot{M}}  \left(\frac{h}{r}\right) 4 \pi \sqrt{-g} dr \simeq -\int_\infty^r L_z\frac{d\Omega}{dr} \sqrt{g_{rr}}dr\;.
 \label{eq:integral}
\end{equation}
To leading order, in the Schwarzschild metric, $L_z\simeq g_{\phi\phi}\Omega$ and
\begin{equation}
    \Omega= \sqrt{\frac{-g_{tt,r}}{g_{\phi\phi,r}}}\;
\end{equation}
(see eq.~[\ref{eq:omega}]). Inserting these expressions into eq.~[\ref{eq:integral}] and performing the integral gives
\begin{equation}
    \int_\infty^r \frac{\Phi}{\dot{M}}  \left(\frac{h}{r}\right) 4 \pi \sqrt{-g} dr \simeq \frac{3GM}{2rc^2}+{\cal O}(r^{-2})\;.
\end{equation}

Similarly, assuming a power-law density profile and evaluating the integral in eq.~(\ref{eq:calV}) gives the same radial dependence but with a different constant coefficient, as was the case in the Newtonian limit (cf eq.~[\ref{eq:kT_Newton}]). Combining all these constant coefficients into one, which we denote by $\zeta$, and inserting the value of the integral into eq.~(\ref{eq:Ti_int}) allows us to write
\begin{equation}
    T_{\rm i}= \frac{m_p c^2}{k_{\rm B}} \frac{R (\hat{\gamma}-1)}{(R+1)} \zeta \left(\frac{GM}{rc^2}\right)\;.
\end{equation}
Note that, near the radius of the photon orbit, the integral in eq.~(\ref{eq:integral}) formally diverges; this is an artifact of the simplifications employed here and is not supported by the temperature profiles found in simulations. For this reason, we will only consider the leading-terms in our analytic model for the heating of ions in the accretion flow, as we have done in the above expression.

\bibliography{shadows.bib}

\end{document}